%% file: main.tex
\documentclass[review]{elsarticle}
\journal{Computerized Medical Imaging and Graphics}

\usepackage{lipsum}
\makeatletter
\def\ps@pprintTitle{%
 \let\@oddhead\@empty
 \let\@evenhead\@empty
 \def\@oddfoot{}%
 \let\@evenfoot\@oddfoot}
\makeatother

\usepackage{amsmath,amssymb,amsfonts}
\usepackage{algorithmic}
\usepackage{graphicx}
\usepackage{textcomp}
\usepackage[export]{adjustbox}
\usepackage{subfig}
\usepackage{lineno,hyperref}
\modulolinenumbers[5]
\usepackage{float}
\usepackage{xcolor}
\usepackage{amsmath}
\UseRawInputEncoding

\begin{document}

    %% Portada
    \begin{frontmatter}
    
        %% Title
        %--------
        \title{WeGleNet: A Weakly-Supervised Convolutional Neural Network for the Semantic Segmentation of Gleason Grades in Prostate Histology Images}
        
        %% Acknowledgements
        \cortext[]{This work was supported by the Spanish Ministry of Economy and Competitiveness through projects DPI2016-77869 and PID2019-105142RB-C21.\\
         \copyright 2021. This manuscript version is made available under the CC-BY-NC-ND 4.0 license http://creativecommons.org/licenses/by-nc-nd/4.0/.}
        
        %% Authors and Afilliations
        \author[aff1]{Julio Silva-Rodr\'iguez}
        \ead{jjsilva@upv.es}
        
        \author[aff2]{Adri\'an Colomer}
        \ead{adcogra@ui3b.upv.es}
        
        \author[aff2]{Valery Naranjo}
        \ead{vnaranjo@dcom.upv.es}
        
        %% Afilliations Descriptions
        \address[aff1]{Institute of Transport and Territory, \textit{Universitat Polit\`ecnica de Val\`encia}, Valencia, Spain}
        
        \address[aff2]{Institute of Research and Innovation in Bioengineering, \textit{Universitat Polit\`ecnica de Val\`encia}, Valencia, Spain}

        %% Abstract
        %-----------
        \input{00_Abstract.tex}
        
        %% Keywords
        %-----------
        \begin{keyword}
        Gleason grading, prostate cancer, semantic segmentation, tissue micro-arrays, weakly supervised.
        \end{keyword}

    \end{frontmatter}

%% Introduction
%-----------
\input{01_Introduction.tex}

%% Related work
%-----------
\input{02_Related_Work.tex}

%% Dataset
%-----------
\input{03_Dataset}

%% Methods
%-----------
\input{04_Methods.tex}

%% Experiments and results
%-----------
\input{05_Experiments_Results.tex}

%% Conclusions
%-----------
\input{06_Conclusions.tex}

%% References
%-------------
\bibliography{references_manual, references_cvblab}

\end{document}

%% file: 00_Abstract.tex
\begin{abstract}

\textit{Background and Objective:}\\
Prostate cancer is one of the main diseases affecting men worldwide. The Gleason scoring system is the primary diagnostic tool for prostate cancer. This is obtained via the visual analysis of cancerous patterns in prostate biopsies performed by expert pathologists, and the aggregation of the main Gleason grades in a combined score. Computer-aided diagnosis systems allow to reduce the workload of pathologists and increase the objectivity. Nevertheless, those require a large number of labeled samples, with pixel-level annotations performed by expert pathologists, to be developed. Recently, efforts have been made in the literature to develop algorithms aiming the direct estimation of the global Gleason score at biopsy/core level with global labels. However, these algorithms do not cover the accurate localization of the Gleason patterns into the tissue. These location maps are the basis to provide a reliable computer-aided diagnosis system to the experts to be used in clinical practice by pathologists.

In this work, we propose a deep-learning-based system able to detect local cancerous patterns in the prostate tissue using only the global-level Gleason score obtained from clinical records during training.

\textit{Methods:}\\
The methodological core of this work is the proposed weakly-supervised-trained convolutional neural network, WeGleNet, based on a multi-class segmentation layer after the feature extraction module, a global-aggregation, and the slicing of the background class for the model loss estimation during training.

\textit{Results:}\\
Using a public dataset of prostate tissue-micro arrays, we obtained a Cohen's quadratic kappa ($\kappa$) of $0.67$ for the pixel-level prediction of cancerous patterns in the validation cohort. We compared the model performance for semantic segmentation of Gleason grades with supervised state-of-the-art architectures in the test cohort. We obtained a pixel-level $\kappa$ of $0.61$ and a macro-averaged f1-score of $0.58$, at the same level as fully-supervised methods. Regarding the estimation of the core-level Gleason score, we obtained a $\kappa$ of $0.76$ and $0.67$ between the model and two different pathologists.

\textit{Conclusions:}\\
WeGleNet is capable of performing the semantic segmentation of Gleason grades similarly to fully-supervised methods without requiring pixel-level annotations. Moreover, the model reached a performance at the same level as inter-pathologist agreement for the global Gleason scoring of the cores.

\end{abstract}

%% file: 01_Introduction.tex
\section{Introduction}
\label{sec:introduction}

\begin{figure}[htb]
\captionsetup[subfloat]{farskip=1pt,captionskip=0.8pt}
    \centering
      \subfloat[\label{fig1a}]{\includegraphics[width=.235\linewidth, frame]{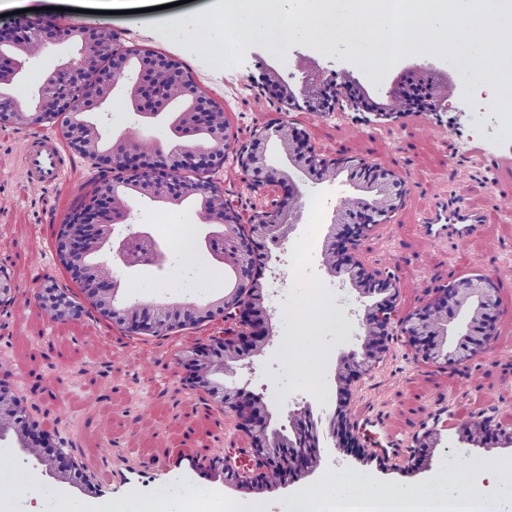}}
      \hspace*{\fill}
      \subfloat[\label{fig1b}]{\includegraphics[width=.235\linewidth, frame]{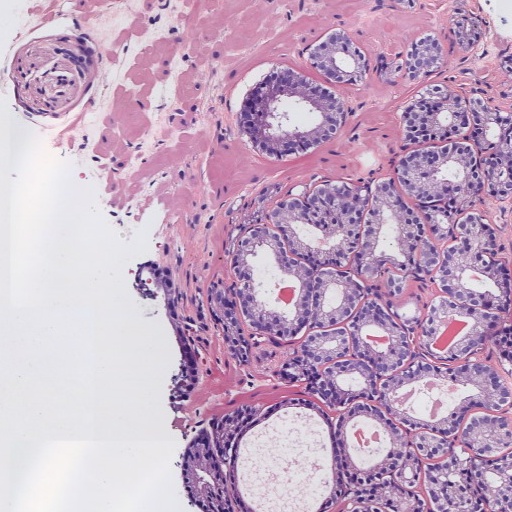}}
      \hspace*{\fill}
      \subfloat[\label{fig1c}]{\includegraphics[width=.235\linewidth, frame]{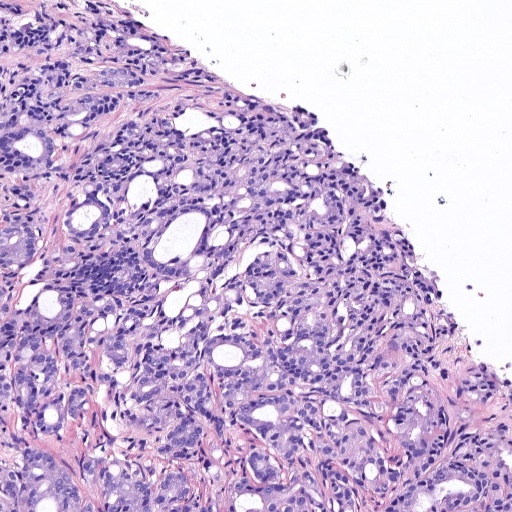}}
      \hspace*{\fill}
      \subfloat[\label{fig1d}]{\includegraphics[width=.235\linewidth, frame]{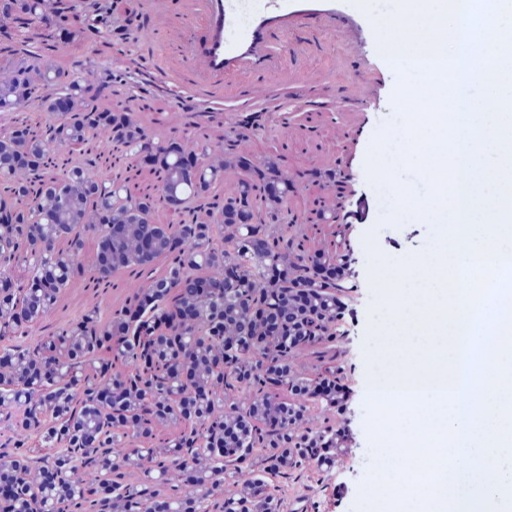}}
    \caption{Histology regions of prostate biopsies. (a): region containing benign glands, (b): region containing GG3 glandular structures, (c): region containing GG4 patterns, (d): region containing GG5 patterns. GG: Gleason grade.}
    \label{fig1}
\end{figure}

% ------------
%% Motivation
Prostate cancer is one of the most common diseases affecting men worldwide. It constitutes $14.5\%$ of all cancers affecting men \cite{wcrf}, and, according to the World Health Organization, the yearly number of new cases will increase by up to $1.8$ million people in this decade \cite{who}. The gold standard for prostate cancer diagnosis and prognosis prediction is the analysis of prostate biopsies under the Gleason grading system \cite{gleason}. This system defines a series of cancerous patterns related to the morphology, distribution, and degree of differentiation of the glands in the tissue. Specifically, in histology slides, the observable Gleason grades (GG) range from $3$ (GG3) to $5$ (GG5). Examples of those patterns are presented in Figure \ref{fig1}.

In clinical practice, small portions of tissue are extracted, laminated, stained with Hematoxylin and Eosin, and finally analyzed under the microscope by expert pathologists using this system. Local cancerous regions of the sample are classified according to the Gleason grades, and finally, the two majority patterns are grouped to obtain a Gleason score as prognosis biomarker (e.g. the Gleason score $5+4=9$ would be assigned to a sample in which the main cancerous Gleason grade is GG5 and the second is GG4). Due to the large size of the biopsies augmented under a microscope, this process results in a high time-consuming and repetitive task, and presents a large intra and inter pathologist variability \cite{Burchardt2008InterobserverMicroarrays}.

In the last decades, the development of digitization devices has allowed the storage of biopsies at microscopic magnifications as digital images. Due to this advance, the field of Computer-Aided Diagnosis (CAD) systems to support pathologists based on computer-vision techniques has experienced a great growth. However, the development of those applications is limited due to the high data-demanding character of deep learning algorithms, and the difficulty in obtaining pixel-level labeled histology images \cite{Komura2018MachineAnalysis}. Normally, pathologists store in the clinical history the global-level diagnosis of the biopsy (e.g. the Gleason score per prostate biopsy). In order to train/build models or develop algorithms able to detect and grade local cancerous patterns, a laborious manual annotation process is required, which must be performed by expert pathologists due to the complexity of the task. In the case of prostate cancer, the different tumor patterns have to be accurately delimited at the pixel level to avoid noisy annotations. Even though multi-resolution graphical user interfaces are provided to clinicians for performing this task, it is a tedious process prone to error. These limitations encourage the development of weakly-supervised deep-learning techniques able to utilize global labels during the training process to accurately identify local cancerous patterns in the images. The main benefit of those methods is that they are not limited to the annotated samples. They can work using histology images labeled only in the global-level patient diagnosis. Recent advances in the literature have proposed the use of the global Gleason score (obtained from the clinical record) to develop CAD systems for biopsy scoring (these works are detailed in Section \ref{ssec:rw2}). Nevertheless, these methods focus on predicting only global biopsy-level markers, while the location of the cancerous structures in the tissue is qualitatively evaluated or not addressed. The classification of local Gleason grades in prostate biopsies is the basis of CAD systems during its use in clinical practice. Accurate heat-maps provide confidence to the pathologists in the daily use of the CAD system, and support the biopsy-level markers provided by the system.

% -----------------------
%% Objective and Novelties

In this work we propose a deep-learning architecture based on convolutional neural networks able to perform a semantic segmentation of the Gleason grades (i.e. non-cancerous tissue, GG3, GG4 or GG5 classes) in prostate histology images, trained via weak supervision using the diagnosed Gleason score of the sample. To the best of the authors' knowledge, this is the first time in the literature that weakly-supervised methods are explored and quantitatively assessed for the local segmentation of cancerous Gleason grades. The main contributions of this research are the following: (i): a weakly-supervised framework based on a convolutional neural network (CNN) architecture able to obtain complementary semantic segmentation maps based on a novel configuration of multi-class activation maps, aggregation layers and the slicing of the background class prediction during training; (ii) the validation of different aggregation layers and regularization techniques to optimize the model; and (iii) the comparison of the proposed weakly-supervised model with fully-supervised state-of-the-art methods.

% -----------------------
%% Organization

The remainder of this paper is divided into five sections. First, Section \ref{sec:relWork} presents an overview of the literature related to this research. Concretely, Section \ref{ssec:rw1} describes the paradigm of weakly-supervised segmentation and the main related computer-vision techniques, Section \ref{ssec:rw2} describes its applications in histology images and Section \ref{ssec:rw3} details the state-of-the-art methods for Gleason grading of local cancerous patterns. Secondly, Section \ref{sec:dataset} describes the database used in the experimental stage. Then, Section \ref{sec:methods} details the methodological core of this work. In particular, Section \ref{ssec:meth1} and \ref{ssec:meth2} describe WeGleNet, our proposed weakly-supervised CNN architecture (contribution (i)), and Section \ref{ssec:meth3} presents the fully-supervised state-of-the-art architectures used as a benchmark to compare our model. Then, Section \ref{ssec:meth4} describes the Gleason scoring method using the class-wise segmentation maps. Section \ref{sec:experiments} describes the experiments carried out in this work. The strategy and figures of merit used in this process are specified in Section \ref{ssec:exp1}. Section \ref{ssec:exp2} presents different ablation experiments related to the optimization process of the proposed architecture (contribution (ii)), and Sections \ref{ssec:exp3} and \ref{ssec:exp4} expose an in-depth validation and comparison of our proposed model against fully-supervised models and previous literature (contribution (iii)). Finally, Section \ref{sec:conclusions} summarizes the main conclusions extracted from our work.

%% file: 02_Related_Work.tex
\section{Related Work}
\label{sec:relWork}

% --------------------------------
%% Weakly-Supervised Learning overview
\subsection{Weakly-Supervised Semantic Segmentation}
\label{ssec:rw1}

Weakly-supervised learning deals with the challenge of using incomplete, scarce, inexact, inaccurate, or noisy information. The problem addressed in this work, image segmentation using just global labels during training, is covered within the Multiple Instance Learning (MIL) scope. MIL works with data clustered on bags of instances, under the assumption that bags labeled as a certain class present, at least, one instance belonging to that class. For one image $X$ composed by the instances (pixels) $x_{ij}$, the bag-level label ($Y$) for a class ($c$) could be interpreted as:

\begin{equation}
    \label{eq:MIL}
       Y_c=
       \begin{cases} 
        1, & \text{if}\ \exists \; x_{ij} : y_c = 1\\ 
        0, & \text{otherwise}
        \end{cases}
\end{equation}
\noindent where $y_c$ is the instance-level label for certain class $c$.

In this topic, two different kinds of classification problems are defined: the prediction of bag-level (global) labels, or the classification of individual instances. In this work, both problems are addressed. A recent extensive review of MIL and its characteristics can be found in \cite{Carbonneau2018MultipleApplications}. Regarding MIL in image classification, convolutional neural networks (CNNs) are the most used technique, since they have demonstrated promising properties for locating objects while performing image-level classification tasks \cite{Oquab2012IsFree, Oquab2014LearningNetworks}.

The approaches to obtain segmentation maps from global-level image classification using CNNs can be divided into aggregation and gradient-based methodologies. Aggregation methods build segmentation maps into the CNN architecture. They are composed of three main blocks: a feature-extraction stage (or base model), an adaptation layer that constructs segmentation maps per class, and a global aggregation layer that resumes each map to one representative value. Then, a multi-label loss function is used to optimize the network weights. The main proposed architectures in this field are WILDCAT \cite{Durand2017WILDCAT:Segmentation} and Attention-MIL \cite{Ilse2018Attention-basedLearning}. WILDCAT constructs the adaptation layer by pooling activation maps after the last convolutional block of the base model and then applies a global-pooling operator to obtain the bag-level probabilities. Attention-MIL joins the adaptation and global aggregation layer by using an attention mechanism that combines all the features obtained in each instance by fully-connected layers. Regarding the gradient methods, the segmentation maps are obtained by post-processing the network output. In this line, the most relevant technique in the literature is the gradient-based class activation maps (Grad-CAMs) \cite{Li2019AnLocalization}. In this technique, the activation maps of the last convolutional block are linearly combined. Each map is weighted by back-propagating gradients in the network from the classification layer, and a ReLU activation is applied to the weights to keep just the features with a positive influence on the classification. Recently, the efforts on weakly supervised semantic segmentation have focused on self-supervised learning. In this methodology, CAMs obtained from gradient-based methods are used as pseudo labels to feed a pixel-level semantic segmentation network. Although these methods have reached promising results, they are still limited by the CAMs used, and the propensity of CNNs to look only at specific and discriminatory patterns. In this line, Ficklenet \cite{Lee2019Ficklenet:Inference} and IRNet \cite{Ahn2019} have proposed the use of center-fixed spatial dropout and class propagation respectively to alleviate this limitation. In all the strategies, the aggregation of the different class-level maps (or CAMs) in a semantic segmentation mask is not straightforward. This process is usually carried out by hand-crafted post-processing. Some methods are based on simply assigning to each pixel the label with the highest probability and let as background those instances with probabilities below certain threshold \cite{Lee2019Ficklenet:Inference}. Other works apply complex fully-connected conditional random fields (CRF) to combine the different class-level maps into one combined mask \cite{Durand2017WILDCAT:Segmentation, Papandreou2015Weakly-andSegmentation, Bency2016WeaklyMaps, Krahenbuhl2011EfficientPotentials}. In our work, we take steps forward in order to solve this limitation, and propose a CNN architecture that obtains complementary multi-class semantic segmentation maps without requiring any post-processing (see Section \ref{ssec:meth1} for further explanation). An extensive survey regarding the application of weakly-supervised learning across different image domains and its current limitations was recently presented in \cite{Chan2020}.

% --------------------------------
%% Weakly-Supervised CNNs in histology overview
\subsection{Weakly-Supervised Segmentation in Histology Images}
\label{ssec:rw2}

Weakly-Supervised learning is a field of increasing interest for histology images, due to the difficulty of preparing large datasets labeled by expert pathologists. While some works just focus on the prediction of bag-level labels in biopsy slides \cite{Campanella2019Clinical-gradeImages, Courtiol2017ClassificationApproach, Arvaniti2018CouplingImages, Wang2019WeaklyAnalysis} carrying out a qualitative evaluation of instance-level (local) classifications, others quantitatively evaluate their proposed models for the local-level classification task \cite{Jia2017ConstrainedSegmentation,Li2016AnProstatectomies, Xu2014WeaklyClassification, Chan2019HistoSegNetImages}. Nevertheless, most of the works only focus on binary classification cancer/no cancer. Early work in \cite{Xu2014WeaklyClassification} proposes a MIL model based on hand-crafted feature extraction (SIFT, color histogram, Local Binary Patterns, etc.), machine learning classifiers and aggregation of the instance-level probabilities for colon cancer detection. Lately, semi-supervised CNNs were used for gland segmentation in prostate images in \cite{Li2016AnProstatectomies}. However, the proposed UNet required to incorporate some instance-level annotations during training to perform properly. Finally, recent work in \cite{Jia2017ConstrainedSegmentation} included previous knowledge by applying constraints in the training stage of a weakly-supervised CNN to control the size of positive instances in the image for colon cancer detection. Recent works have used weakly-supervised CNNs approaches for multi-class semantic segmentation. Concretely, HistoSegNet, introduced in \cite{Chan2019HistoSegNetImages}, performs a weakly-supervised segmentation of different tissue types in histology images based on CNNs and Grad-CAM gradient method. Then, a complex hand-crafted post-processing is proposed to join the class-level segmentation maps and to include the background class. 

% --------------------------------
%% Weakly-Supervised CNNs in histology overview
\subsection{Prostate Gleason Grading}
\label{ssec:rw3}

In the analysis of prostate histology samples, as mentioned previously, there are two main tasks: the grading of local structures using the Gleason system, and the global scoring.

First works in this field focused on fine-tuning well-known CNN architectures in a supervised patch-level classification, with the requirement of pixel-wise expert annotations. In this line, Nir et al. \cite{Nir2018AutomaticExperts,Nir2019ComparisonImages} obtained a patch-level Cohen's quadratic kappa ($\kappa$) of $0.60$ in the validation set, while $0.55$ and $0.49$ was reached by Arvaniti et al. in \cite{Arvaniti2018AutomatedLearning} in the test cohort referenced to two different pathologists. Then, the percentage of each cancerous tissue in the sample was calculated from the patch-level probabilities to predict the Gleason score of the sample. Arvaniti et al. \cite{Arvaniti2018AutomatedLearning} obtained with this method a $\kappa$ of $0.76$ and $0.71$ against the annotations of two different pathologist, at the level of the inter-pathologist agreement ($\kappa$ = $0.71$).

Latest works in the literature have started to develop weakly-supervised techniques to avoid the tedious process of pixel-level labeling of Gleason grades. These techniques are based on assigning the global labels (i.e. the primary and secondary grades obtained from the Gleason score) to patch-level regions of interest (i.e. glandular or nuclei structures). Then, convolutional neural networks are trained to perform a patch-level classification with the obtained pseudo-ground truth. The selection of regions of interest in the tissue are based on different approaches, detailed in the following lines. The work in \cite{Bulten2020AutomatedStudy} developed a semi-supervised pipeline detecting the glandular tissue via a UNet trained with manual annotations. A few works works focus on selecting these regions with larger amounts of nuclei, based on color \cite{JimenezdelToro2017ConvolutionalScore, Otalora2019StainingPathology} or Laplacian filters \cite{Strom2020ArtificialStudy}. Finally, the work in \cite{Campanella2019Clinical-gradeImages} directly assigns the global label (cancerous against non cancerous) to all the patches in the tissue. All previous methods train patch-level convolutional neural networks with the obtained pseudo-ground truth, and finally they combine the patch-level predictions to obtain the global score. The first works aggregate the predictions using the percentage of each Gleason grade in the sample and then they train different machine learning models to predict the global Gleason score \cite{Bulten2020AutomatedStudy, JimenezdelToro2017ConvolutionalScore,Strom2020ArtificialStudy}. Also, novel approaches combine the patches using the features extracted by the CNN through recurrent neural networks \cite{Campanella2019Clinical-gradeImages}. Although the aforementioned methods provide promising results for Gleason scoring of prostate biopsies, the assumptions made to develop their weakly-supervised pipeline could be affecting the local grading of cancerous patterns. To the best of the authors' knowledge, none of previous works in the literature focus on locating the Gleason grades in the tissue using weakly-supervised learning. They only perform a qualitative evaluation of the heat-maps obtained by their models.

%% file: 03_Dataset.tex
\section{Dataset}
\label{sec:dataset}

The experiments described in this work were carried out using the public dataset presented by Arvaniti et al. in \cite{Arvaniti2018AutomatedLearning}\footnote{We contacted the corresponding authors to obtain the dataset.}. This dataset consists of $886$ prostate Tissue Micro-Arrays (TMAs, samples of representative regions of cancerous biopsies known as cores), digitized at $40\times$ magnification in images of size $3100^2$ pixels. The cores include pixel-level annotations of Gleason grades and benign structures, and global labels of Gleason scores (primary and secondary Gleason grades in the sample). The distribution of the Gleason grades (GG) in the cores is distributed as follows: $421$, $387$ and $148$ cores with GG3, GG4 and GG5, respectively. Regarding the pixel-level annotations, the dataset includes five different classes: benign tissue, GG3, GG4, GG5, and background. In order to evaluate our proposed methodology, the benign and background classes are joined in the non-cancerous (NC) class. To establish fair comparisons with previous literature, the partition of the dataset proposed by Arvaniti et al. was used for training, validating, and testing. Note that the test cohort contains pixel-level annotations made by two different expert pathologists.

%% file: 04_Methods.tex
\section{Methods}
\label{sec:methods}

% --------------------------------
%% Multi-Class Semantic Segmentation Architecture

\subsection{WeGleNet: Weak-Supervised Gleason Grading Network}
\label{ssec:meth1}

\begin{figure*}[htb]
    \begin{center}
    \includegraphics[width=1\textwidth]{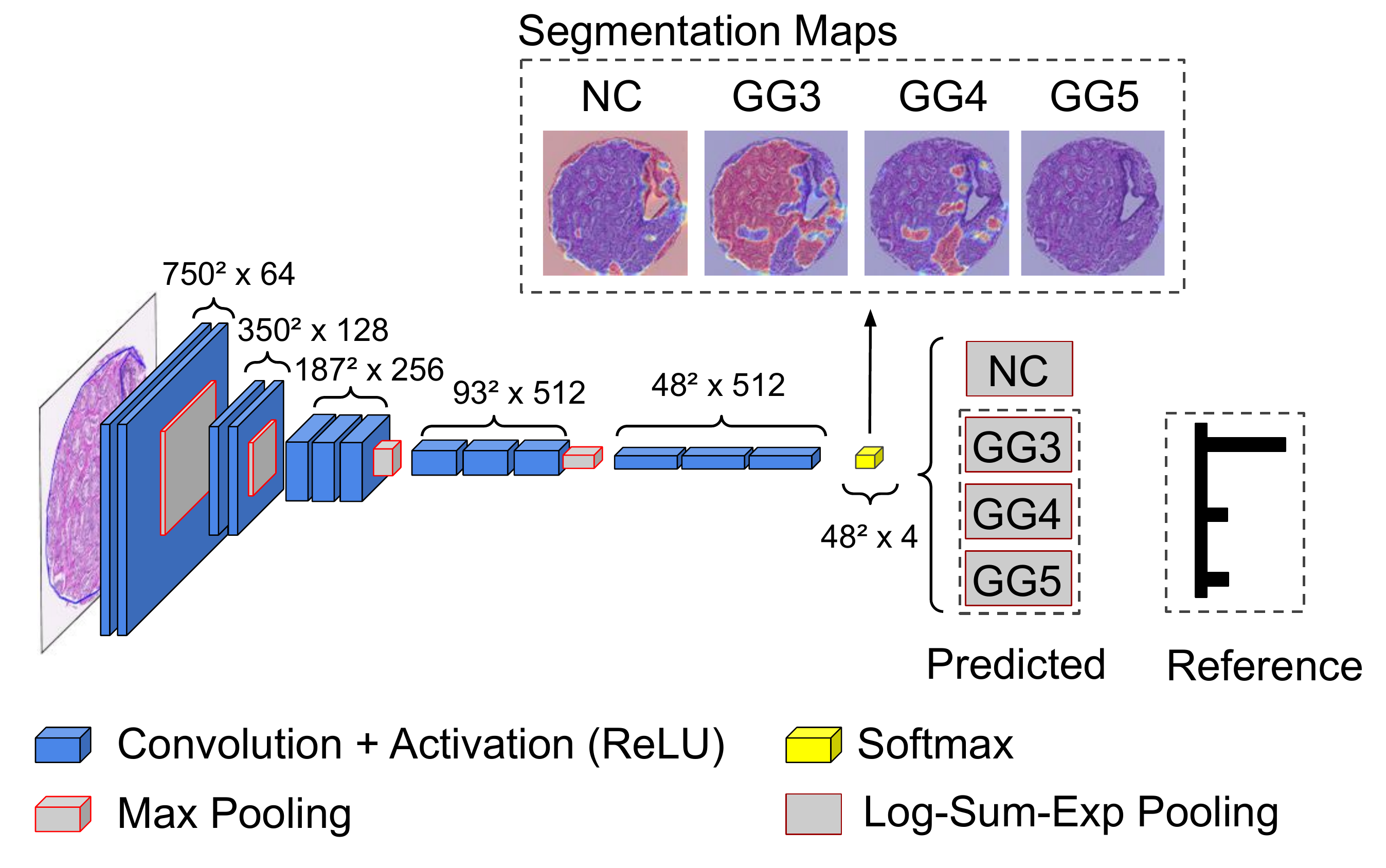}
    \caption{WeGleNet, weakly-supervised framework for semantic segmentation of local cancerous patterns via Gleason grading using the Gleason score of the global sample during the training stage. NC: non cancerous; GG3: Gleason grade $3$; GG4: Gleason grade $4$; GG5: Gleason grade $5$.}
    \label{fig2}
    \end{center}
\end{figure*}

The methodological core of this work consists of a convolutional neural network able to predict semantic segmentation maps of non-cancerous, Gleason grade $3$ (GG3), GG4, and GG5 tissue in prostate histology images, trained using global labels of the grades present in the tissue during training. The proposed weak-supervised Gleason grading network (WeGleNet) is presented in Figure \ref{fig2}.

The architecture is composed of three main components: the base model, the segmentation (also called adaptation) layer, and the global-aggregation operation, and it takes as input the prostate core image, which is resized to $750^2$ pixels due to computational limitations. First, the base model is in charge of extracting automatic-learned features from the input image. Concretely, the VGG19 architecture \cite{Simonyan2014VeryRecognition} is used. This is based on convolutional blocks with an increasing number of filters with $3\times3$ kernels with ReLU activation and dimensional reduction via max pooling of size $2\times2$. In order to reduce the over-fitting during the training stage, weights are initialized using the VGG19 model pretrained in the ImageNet dataset \cite{Deng2009ImageNet:Database}. Secondly, the segmentation layer applies to the output convolutional feature volume of the base model as many convolutional filters of size $1\times1$ as classes to be predicted. This layer also computes a softmax activation along the class dimension generating a multi-class segmentation volume of activation maps, in which each value represents the probability of that pixel of belonging to a class. During the inference stage, this layer will be the model output, and each segmentation map will be resized to the original core dimensions ($3100^2$ pixels). During the training stage, the pixel-level probabilities in the activation maps are aggregated in order to output one global probability per class ranging between $0$ and $1$. This operation is performed by a global-aggregation layer, which is detailed in Section \ref{ssec:meth2}. This aggregation of instance-level predictions embedded in the training stage of the model avoids previous assumptions in the literature to locate the regions of interest in the tissue. Then, binary cross-entropy is used as a loss function. As all cores contain non-cancerous regions, the loss function is only calculated using the Gleason grade classes (i.e. GG3, GG4, and GG5). Thus, the NC class segmentation map gathers those patterns not related to cancer but does not contribute to the calculation of the loss function. This strategy allows obtaining complementary segmentation maps including the background class (in our case non-cancerous class). This is a step forward compared to previous methods, which were based on the individual prediction of segmentation maps per class, and complex post-processing to join them including the background class (see Section \ref{ssec:rw2} for a more detailed explanation of these methods).

During the training stage, two techniques are carried out to regularize the model and avoid over-fitting: data augmentation and hide-and-seek \cite{Singh2017Hide-and-Seek:Localization}. Data augmentation is performed by transforming the input images with random translations, rotations and mirroring in each iteration. Hide-and-seek (HS) is a method that regularizes weakly-supervised-trained architectures by replacing random patches of the images with the average intensity level of the input. In each iteration, the hidden patches vary, and thus the network is forced to focus on heterogeneous patterns during training. The input image is divided into patches of $75^2$ pixels, which have a $25\%$ probability of being hidden in each iteration.

% --------------------------------
%% Aggregation Layers

\subsection{Global-Aggregation Layers}
\label{ssec:meth2}

Global-aggregation layers summarize information from all spatial locations in the activation maps ($x_{ij}$) to one representative value ($p$). For this task, we propose the use of the log-sum-exponential (LSE) layer \cite{Pinheiro2015FromNetworks} in WeGleNet, which is defined as:

\begin{equation}
    \label{eq:lse}
    p_{LSE} = \frac{1}{r} \cdot log\left[\frac{1}{S}\cdot\sum_{(i,j)\in S}^{} exp(r\cdot x_{ij})\right]
\end{equation}
    
\noindent where $S$ constitutes the number of pixels in the activation map $x_{ij}$ and $r$ is a parameter to be optimized.

The LSE operation permits us to obtain a domain-specific representation of the activation map via the parameter $r$, with large values of $r$ ($r\to\infty$) similar to a global-max pooling operation (GMP) \cite{Oquab2014LearningNetworks} and small values ($r\to0$) equivalent to a global-average operation (GAP) \cite{Lin2014NetworkNetwork}. The $r$ parameter is empirically fixed by optimizing the model performance in the validation cohort (see Section \ref{ssec:exp2}). By this procedure, the training stage overcomes the limitations of the other global-aggregation layers (i.e. GAP assumes that the pattern is uniformly distributed across with the activation map, and GMP could produce over-fitting to small, specific patterns).

% --------------------------------
%% Supervised CNNs

\subsection{Fully-Supervised CNNs}
\label{ssec:meth3}

To compare our proposed weakly-supervised framework, two state-of-the-art supervised architectures for semantic segmentation of Gleason grades are implemented. To take advantage of the pixel-level annotations, patches are extracted from the cores with a size of $750^2$ pixels and a step of $350$. Due to hardware limitations during training, patches are resized to $224^2$ pixels. Then, a UNet architecture and a classifier based on a patch-level VGG19 fine-tuned network (VGG19Sup) are selected as supervised architectures to be compared to the WeGleNet model. It is important to highlight that these methods require an accurate pixel-level labeling of the images. The implementation of the models is detailed in the following lines.

VGG19Sup is based on training a patch-level multi-class classifier and then modifying the architecture to obtain segmentation maps. VGG19Sup is composed of a feature-extraction stage using VGG19 backbone pre-trained in Imagenet dataset, a global-average pooling (GAP) to aggregate the activation maps, and a fully-connected layer with as many neurons as classes to predict and soft-max activation as output. In this method, each patch is labeled as the majority grade annotated. If none Gleason grade is annotated, the patch is labeled as non-cancerous. Training is performed by optimizing the categorical cross-entropy as loss function. For the inference of segmentation maps, the output fully-connected layer is converted in a convolutional layer with kernel $1\times1$, which is applied over the activation volume previous to the GAP layer to obtain a segmentation map per class. This approach is equivalent to using a class activation map (CAM) post-processing, but the segmentation maps are obtained directly from the CNN in an end-to-end manner. This method was previously used by Arvaniti et al. in \cite{Arvaniti2018AutomatedLearning} to obtain the probability maps in prostate samples.

Regarding the UNet architecture \cite{Ronneberger2015U-net:Segmentation}, it is based on a symmetric encoder-decoder path. In the encoder, feature extraction is carried out based on convolutional blocks and dimensional reduction through max-pooling layers. Each convolutional block increases the number of filters by a factor of $2\times$, starting from $64$ filters up to $1024$. After each block, the max-pooling operation reduces the dimensions of the activation maps in a factor of $2\times$. Then, the decoder path builds the segmentation maps, recovering the original dimensions of the image. The reconstruction process is based on deconvolutional layers with filters of size $3\times3$ and ReLU activation. These layers increase the spatial dimensions of the activation volume in a factor of $2\times$ while reducing the number of filters by a half. Then, the encoder features from a specific level are joined with the resulting activation maps of the same decoder level by a concatenation operation, feeding a convolutional block that combines them. The convolutional block used during both encoder and decoder paths includes residual connections \cite{He2016DeepRecognition} to improve the model optimization. This residual UNet configuration was proposed in \cite{Zhang2018RoadU-Net}, and showed to outperform other configurations for Gleason grading in \cite{Kalapahar2020GleasonU}. It consists of three convolutional layers with $3\times3$ kernels and ReLU activation. The output of the last convolutional layer of the block in connected via a shortcut residual operation with the output of the first layer. Finally, after the decoder, a $1\times1$ convolutional layer creates the segmentation probability maps. The loss function used during the training process is the categorical $Dice$ used in \cite{Kalapahar2020GleasonU}.

During the inference stage, the supervised models are used to predict the entire core instead of local patches. Cores are resized to match the resolution used during training, and then the output segmentation maps are resized to the original dimensions of the cores ($3100^2$ pixels). 

% --------------------------------
%% Gleason scoring

\subsection{Global Gleason Scoring}
\label{ssec:meth4}

Once the probability maps per class are obtained, the Gleason score of the sample is inferred from the percentage of each class $k$ in the tissue, $w^k$. In \cite{Arvaniti2018AutomatedLearning}, the Gleason score is obtained assigning the majority and secondary grades in terms of percentage, considering only the classes above certain threshold $c$. In this work, we introduce another term, $d$, which models the tendency of pathologists to focus on the majority cancerous pattern if it is widespread in the tissue. Thus, the final percentage weights are assigned to each class such that:

\begin{equation}
    \label{eq:agg}
       w^k=
       \begin{cases} 
        0, &  \text{if}\ \quad \max_{k'} \; w^{k'} > d \quad \text{and} \quad k \neq argmax_{k'} \; w^{k'}
        \\ 
        w^k, & \text{otherwise}
        \end{cases}
\end{equation}
\noindent where $k$ denotes the different classes, i.e. Gleason grade $3$, $4$ and $5$.

The operator $d$ adapts the weakly-supervised framework to the global scoring procedure in clinical practice. Pathologists annotate regions focusing on primary patterns, while the weakly supervised model performs a more fine-grained segmentation, that increases the percentage of secondary patterns. Thus, $d$ allows to suppress the system's confidence on these patterns for the global scoring task. The values of the parameters $c$ and $d$ are empirically fixed in the validation set to optimize the results.

%% file: 05_Experiments_Results.tex
\section{Experiments and Results}
\label{sec:experiments}

% --------------------------------
%% Multi-Class Semantic Segmentation Architecture

\subsection{Experimental Strategy and Metrics}
\label{ssec:exp1}

\begin{figure}[htb]
\begin{center}
\includegraphics[width=.8\textwidth]{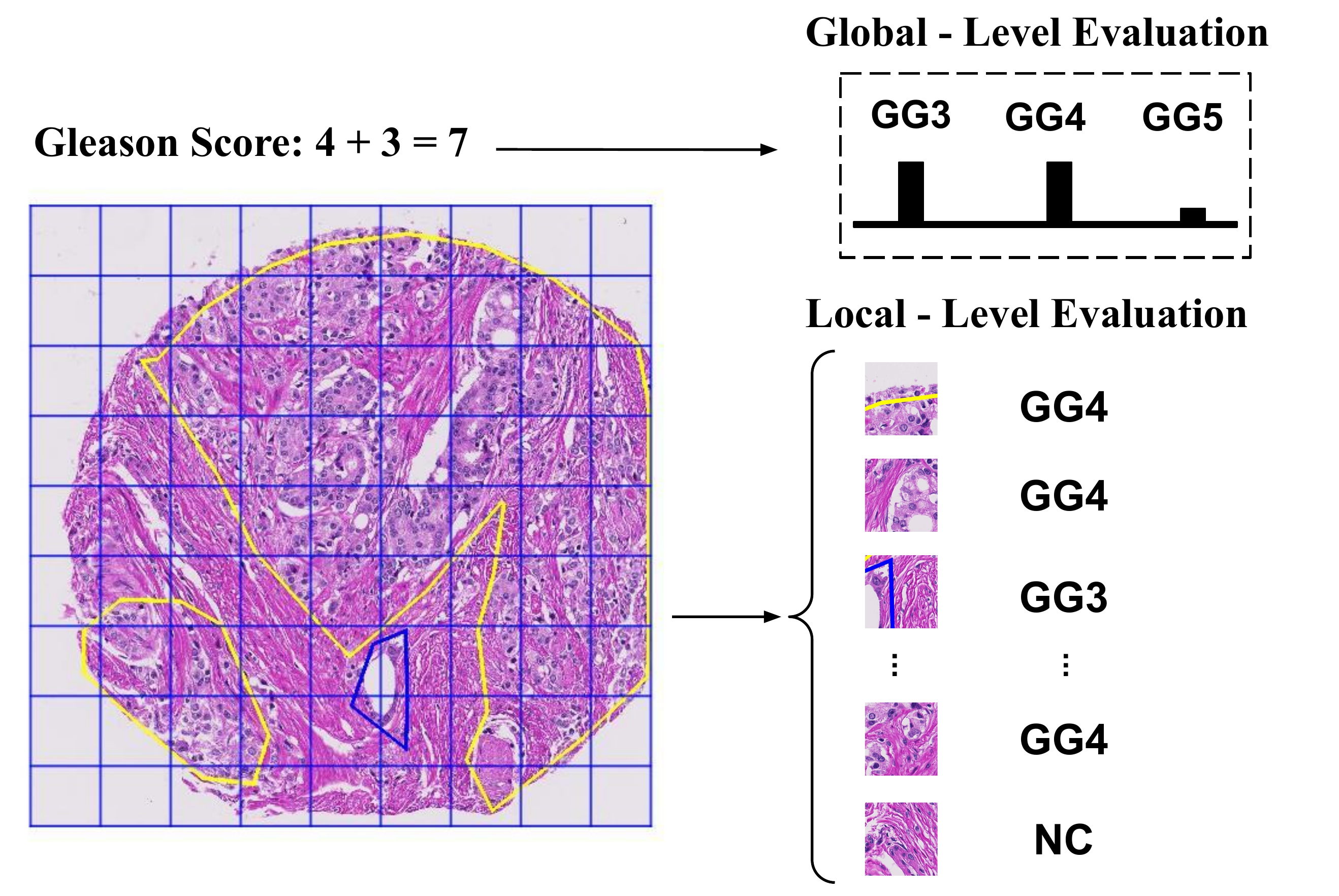}
\caption{Strategies for the evaluation of the model performance. NC: non cancerous; GG: Gleason grade. The core-level (global) predictions are evaluated using the Gleason score. The local-level predictions are evaluated at pixel-level or using small patches extracted from the core.}
\label{fig:strategy}
\end{center}
\end{figure}

In order to validate the proposed WeGleNet model, two types of figures of merit are extracted from the model output: global-level (bag-level in the MIL framework) and local-level (instance-level) metrics. Figure \ref{fig:strategy} illustrates the evaluation strategy.

Global-level metrics are obtained comparing the multi-label prediction of the WeGleNet in the global-aggregation layer and the Gleason grades observed in the core using the reference Gleason score. This evaluation is used to optimize the weakly-supervised model using the Area Under ROC curve (AUC) as a figure of merit. The decision of using this metric during the optimization stage is related to being closer to the output probabilities of the model. Finally, during the comparison of the model performance with previous literature (Section \ref{ssec:exp3}) the Cohen's quadratic kappa ($\kappa$) \cite{Cohen1968WeightedCredit} is obtained for the Gleason score prediction. This agreement statistic takes into account that in a set of ordered classes, errors between adjacent classes should be less penalized.  

Regarding the local-level evaluation, this is performed to analyze the capability of the trained model for segmenting the Gleason grades in the tissue. During WeGleNet optimization and its comparison with fully-supervised methods for semantic segmentation, metrics are obtained at pixel level. The obtained figures of merit are the accuracy (ACC), f1-score per class, the macro-average (F1), mean intersection over union (mIoU) and Cohen's quadratic kappa ($\kappa)$. Usually, in the Gleason grading literature, the local grading of cancerous patterns is evaluated at patch level to avoid underestimation of the model performance due to an inaccurate pixel-level annotation in the ground truth. Therefore, WegleNet is evaluated at patch level for the comparison of its performance with previous state-of-the-art works in this field. In order to establish fair comparisons with previous results reported in the literature in the used dataset, patch-level labels are obtained as proposed by Arvaniti et al. \cite{Arvaniti2018AutomatedLearning}. Concretely, patches are extracted using a moving-window of size $750^2$ and a step of $350$ pixels. Patches with multiple or no annotations were discarded, and the remaining were labeled by majority voting according to the annotations in the central region of the patch (i.e. benign, GG3, GG4 or GG5).

The remainder part of the experimental section describes the experiments carried out to optimize the WeGleNet architecture (Section \ref{ssec:exp2}), and its comparison on the local-level segmentation of Gleason grades with supervised methods (Section \ref{ssec:exp3}) and with previous works using the same dataset (Section \ref{ssec:exp4}).

% --------------------------------
%% Model optimization

\subsection{Model Optimization}
\label{ssec:exp2}

In the first experiments, the objective was to optimize the WeGleNet architecture for semantic segmentation using global-level labels (i.e. the presence of certain Gleason grade in the core). The model performance was studied under the different regularization techniques and global-aggregation layers. WeGleNet model was trained using the proposed log-sum-exponential (LSE), global-max (GMP) and global-average (GAP) pooling. In LSE layer, different values of the $r$ parameter, $r = \{1, 5, 8, 10, 15, 25\}$, were used. In addition, to compare the performance of the LSE with respect to an automatic-learned combination of GMP and GAP, a mixed-pooling (MixP) aggregation layer is implemented such that:

\begin{equation}
    \label{eq:alpha}
    p_{MixP} = \alpha\cdot p_{GMP} + (1-\alpha)\cdot p_{GAP}
\end{equation}
\noindent where $\alpha$ is a parameter learned during training.

\begin{figure}[htb]
    \begin{center}
    %\captionsetup[subfloat]{farskip=1pt,captionskip=0.8pt}
          \qquad\qquad
          \subfloat[\label{fig:res1a}]{\includegraphics[width=.8\linewidth]{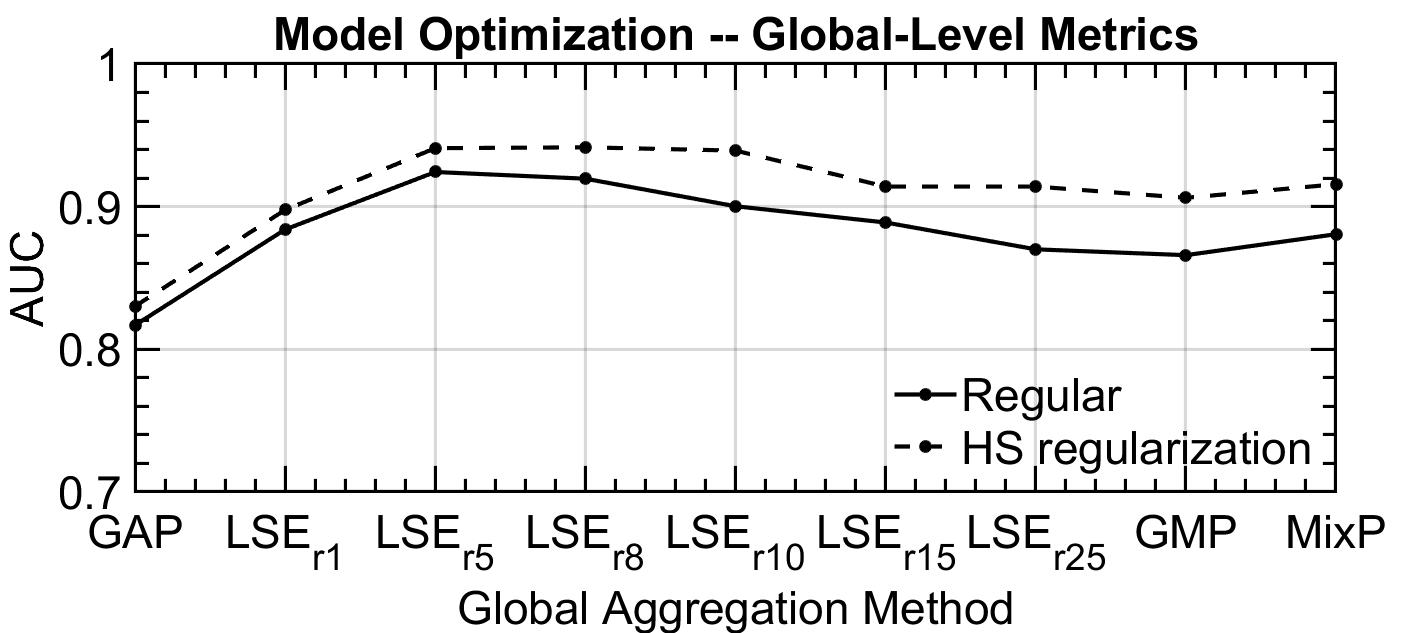}}
          \hspace*{\fill}
    
          \qquad\qquad
          \subfloat[\label{fig:res1b}]{\includegraphics[width=.8\linewidth]{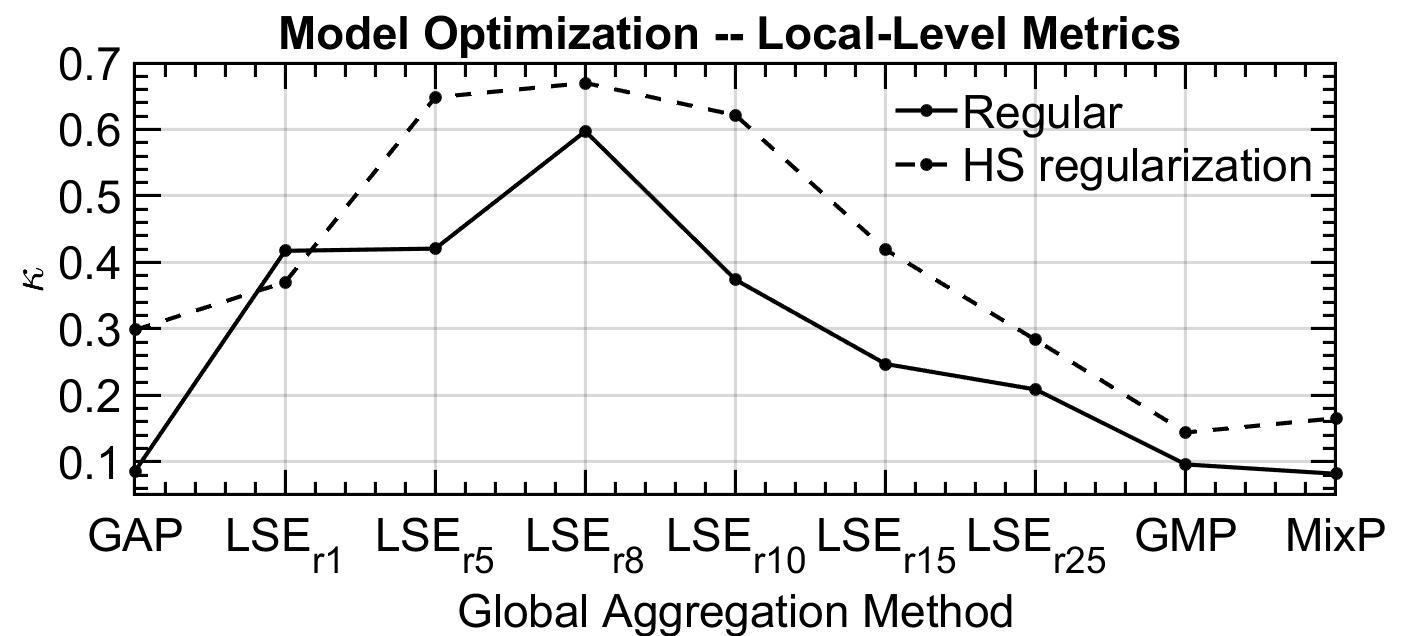}}
          \hspace*{\fill}
    
        \caption{Model performance using different global-aggregation methods and regularization techniques. (a): global prediction performance; (b): pixel-level segmentation performance. HS: hide-and-seek; GAP: global-average pooling; LSE: log-sum-exponential pooling; GMP: global-max pooling; MixP: mixed pooling.}
    
        \label{fig:res1}
    \end{center}
\end{figure}

The use of hide-and-seek (HS) regularization was validated by training the models with and without it. The training was performed in mini-batches of $8$ images, and Stochastic Gradient Descent (SGD) was used as the optimizer with a learning rate of $1\cdot10^{-3}$. Exponential decay in the learning rate was applied in the last 20 epochs to stabilize the model weights such that: $\eta = 1\cdot10^{-3}\cdot e^{-0.1\cdot t}$, where $\eta$ is the applied learning rate and $t$ is the epoch. The training was carried out during $120$ epochs, which were increased to $400$ when applying HS regularization. WeGleNet was trained using the training cohort, and early stopping was applied by keeping the weights of the model obtaining the best performance in the validation set (in terms of the obtained losses). After each experiment, segmentation maps were obtained from the segmentation layer, and core-level predictions were obtained from the global-aggregation layer using the images of the validation cohort. The scripts to reproduce the experiments reported in this work are publicly available on (\url{https://github.com/cvblab/prostate_wsss_weglenet}). Figures of merit related to global-level predictions and pixel-level segmentation are presented in Figure \ref{fig:res1} (a) and (b) respectively.

Regarding the obtained results, LSE pooling showed superior performance compared to other global-aggregation techniques. In particular, the best results were obtained using $r=8$ ($LSE_{r8}$), with an AUC of $0.9243$ for the core-level detection of Gleason grades and a $\kappa$ of $0.5973$ for the pixel-level segmentation. Hide-and-Seek regularization (HS) showed to improve the results in all the experiments, forcing the model to focus on all the patterns of the images. Thus, results improved to an AUC of $0.9416$ and a $\kappa$ of $0.6699$ in the best-performing model, WeGleNet - $LSE_{r8}$. Finally, a high correlation was observed between the global-level and the local-level performance of the model. A Pearson correlation coefficient of $0.5462$ was obtained between $\kappa$ and AUC when using HS regularization. Then, improvements in the global-level predictions produced a better segmentation of the Gleason grades. This promising behavior indicates that the model can be optimized without any pixel-level annotations.   

% --------------------------------
%% Weak vs Strong supervision

\subsection{Weak Supervision vs. Strong Supervision}
\label{ssec:exp3}

\begin{figure}
    \centering
    
      \subfloat{\includegraphics[width=.19\linewidth]{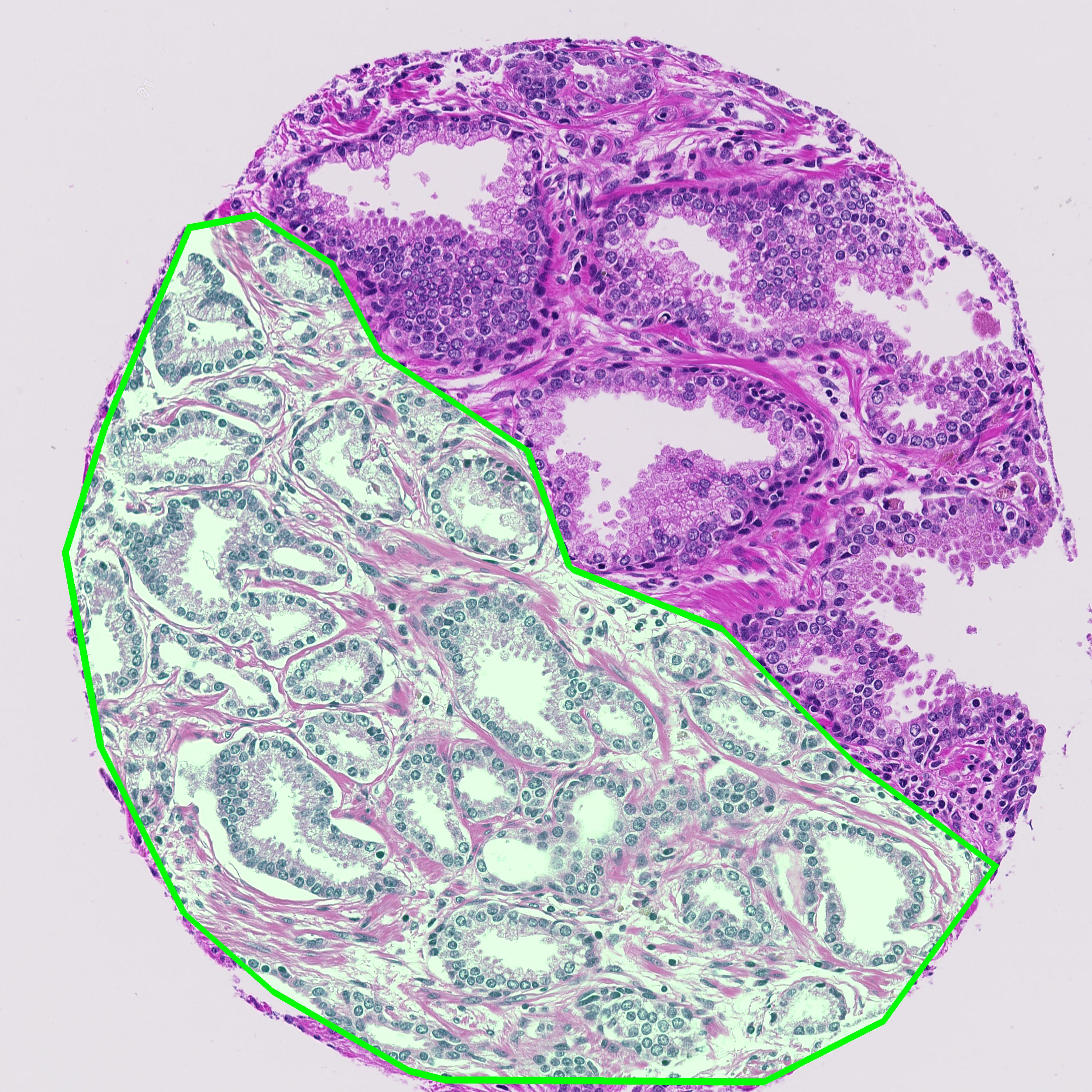}}
      \hspace*{\fill}
      \subfloat{\includegraphics[width=.19\linewidth]{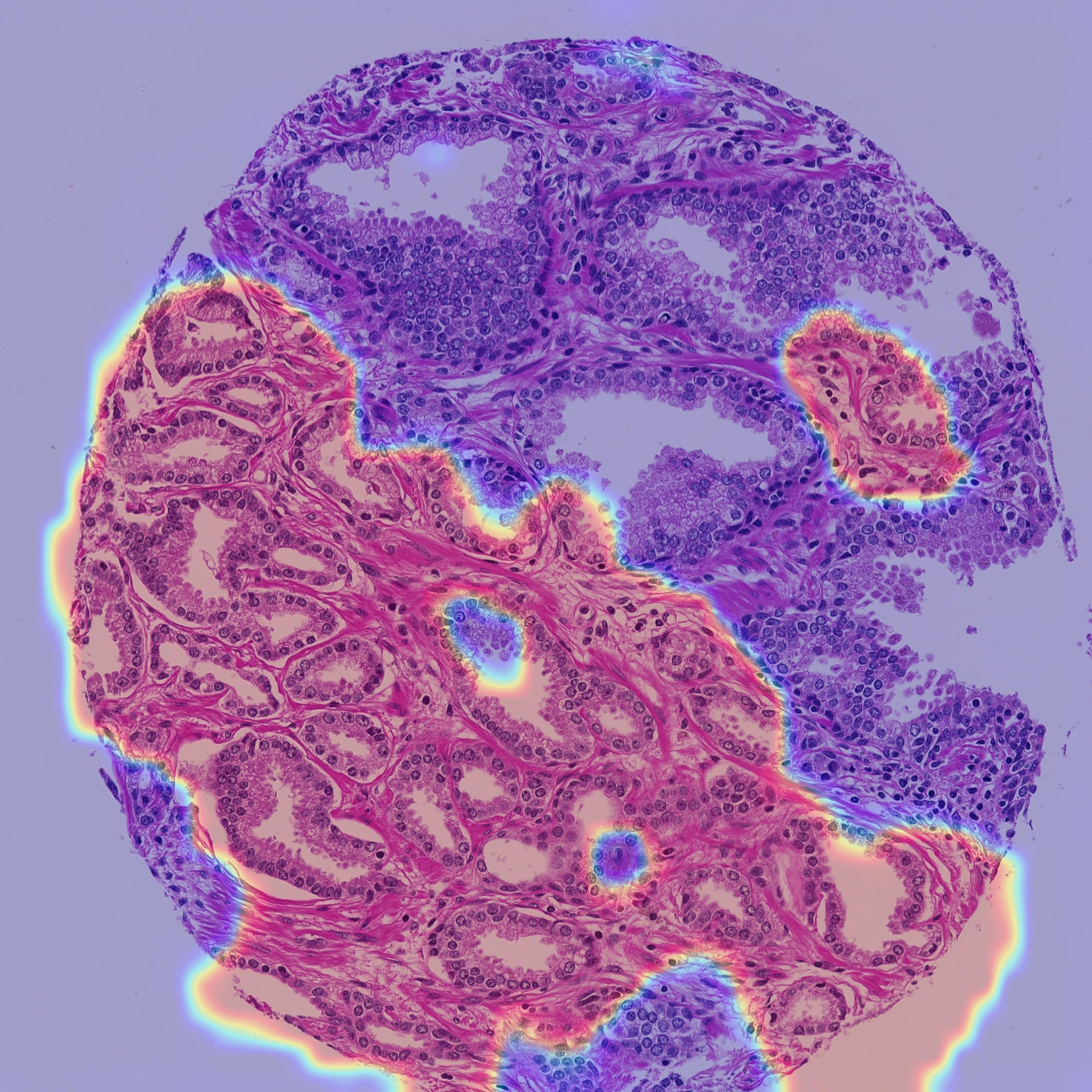}}
      \hspace*{\fill}
      \subfloat{\includegraphics[width=.19\linewidth]{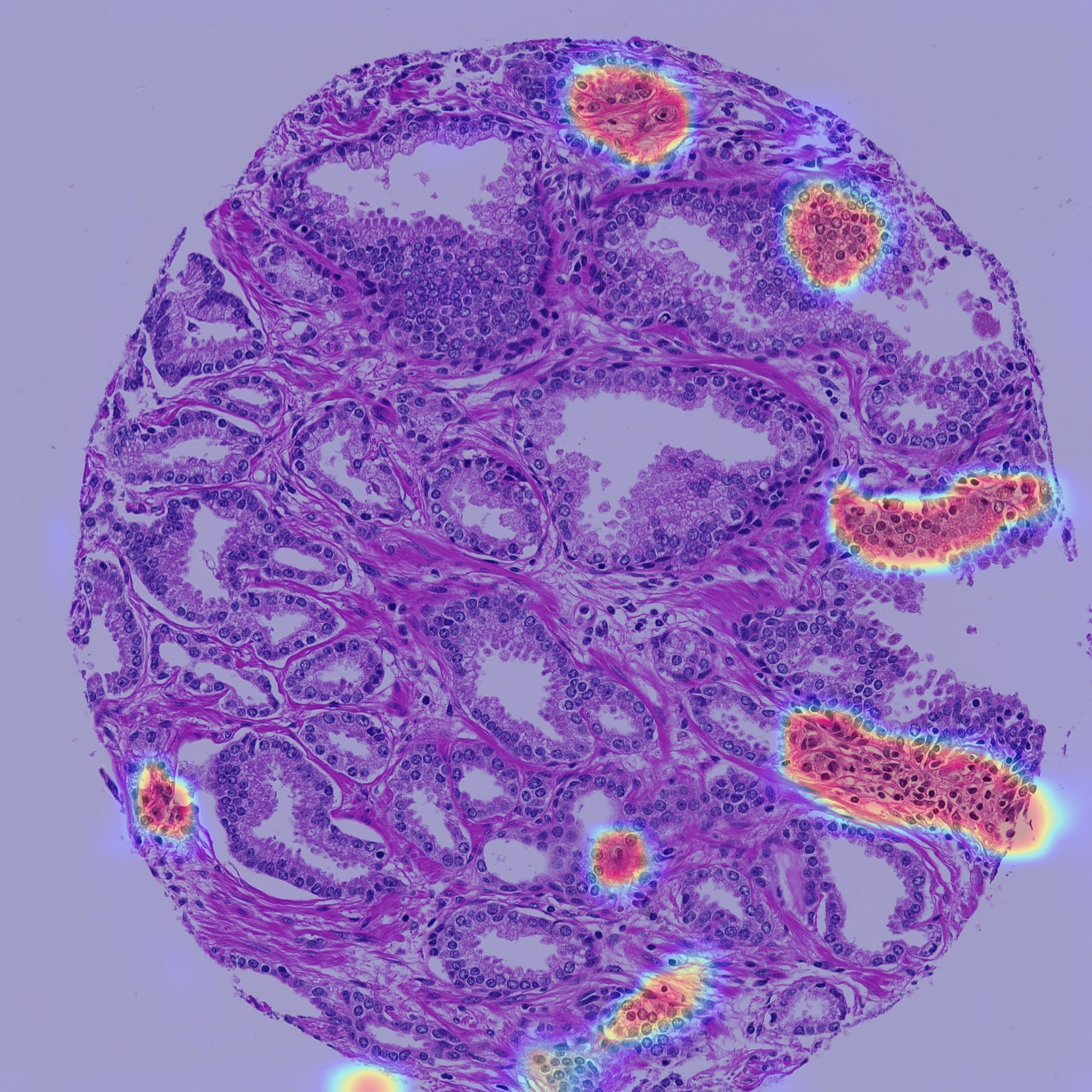}}
      \hspace*{\fill}
      \subfloat{\includegraphics[width=.19\linewidth]{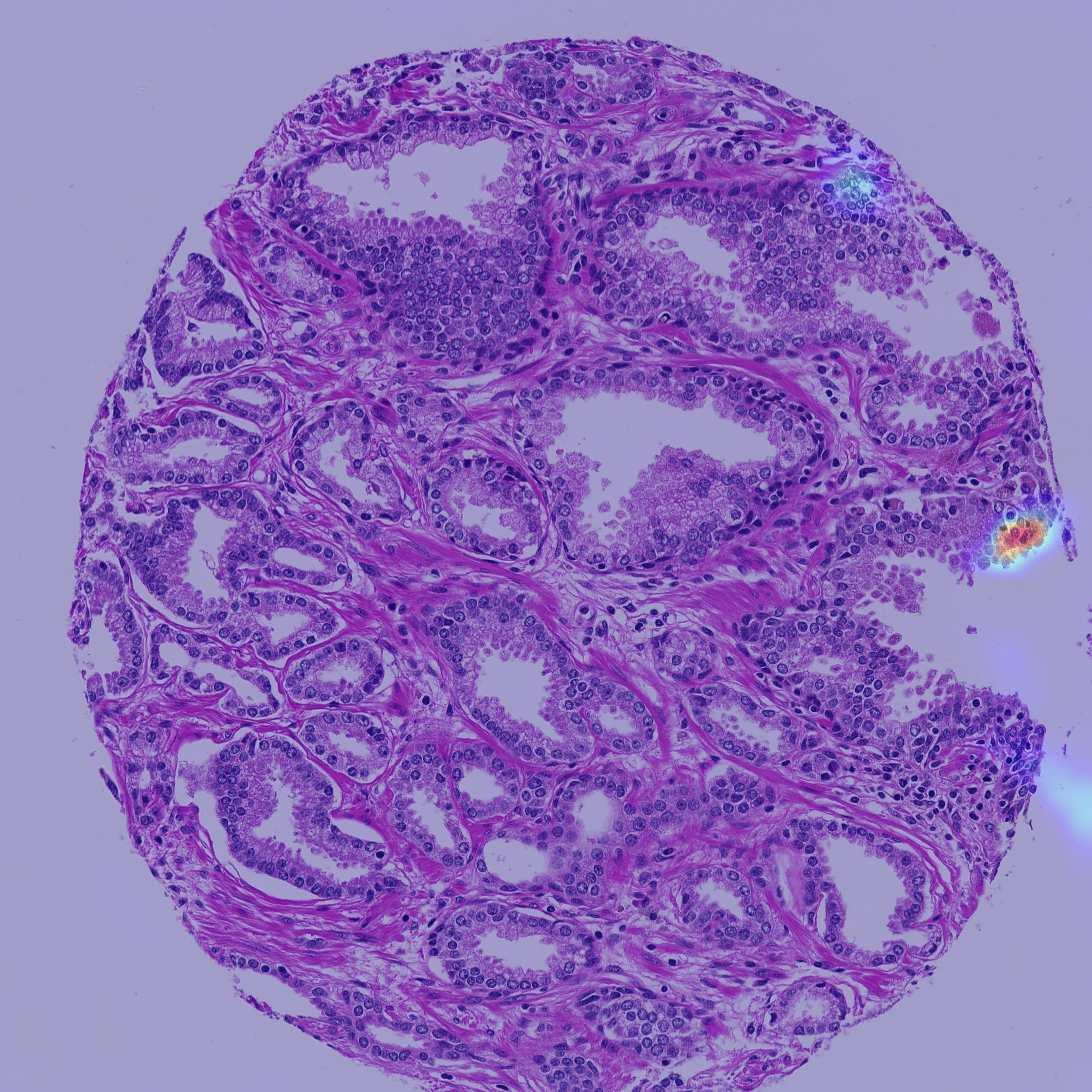}}
      \hspace*{\fill}
      \subfloat{\includegraphics[width=.19\linewidth]{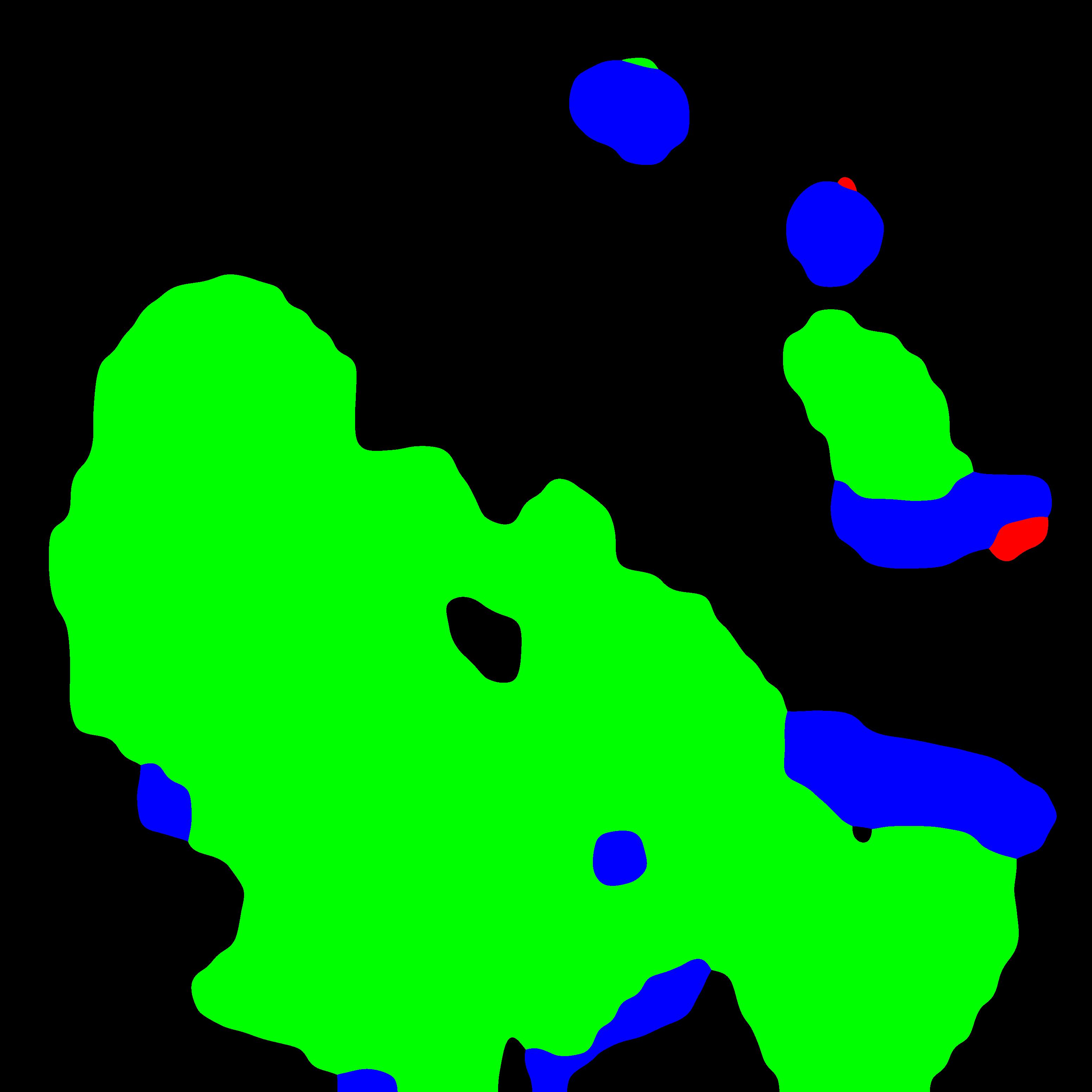}}
      \hspace*{\fill}    
    
      \subfloat{\includegraphics[width=.19\linewidth]{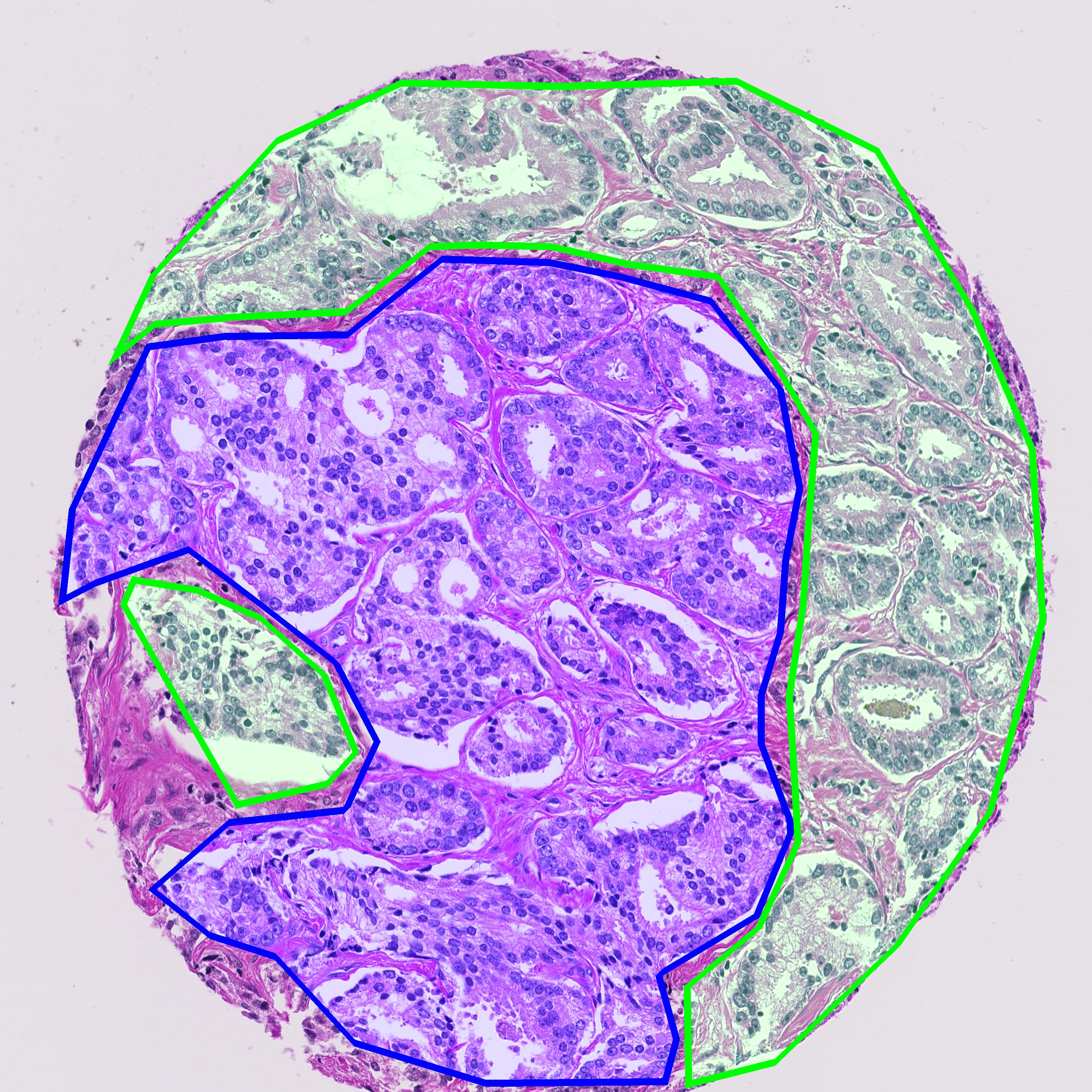}}
      \hspace*{\fill}
      \subfloat{\includegraphics[width=.19\linewidth]{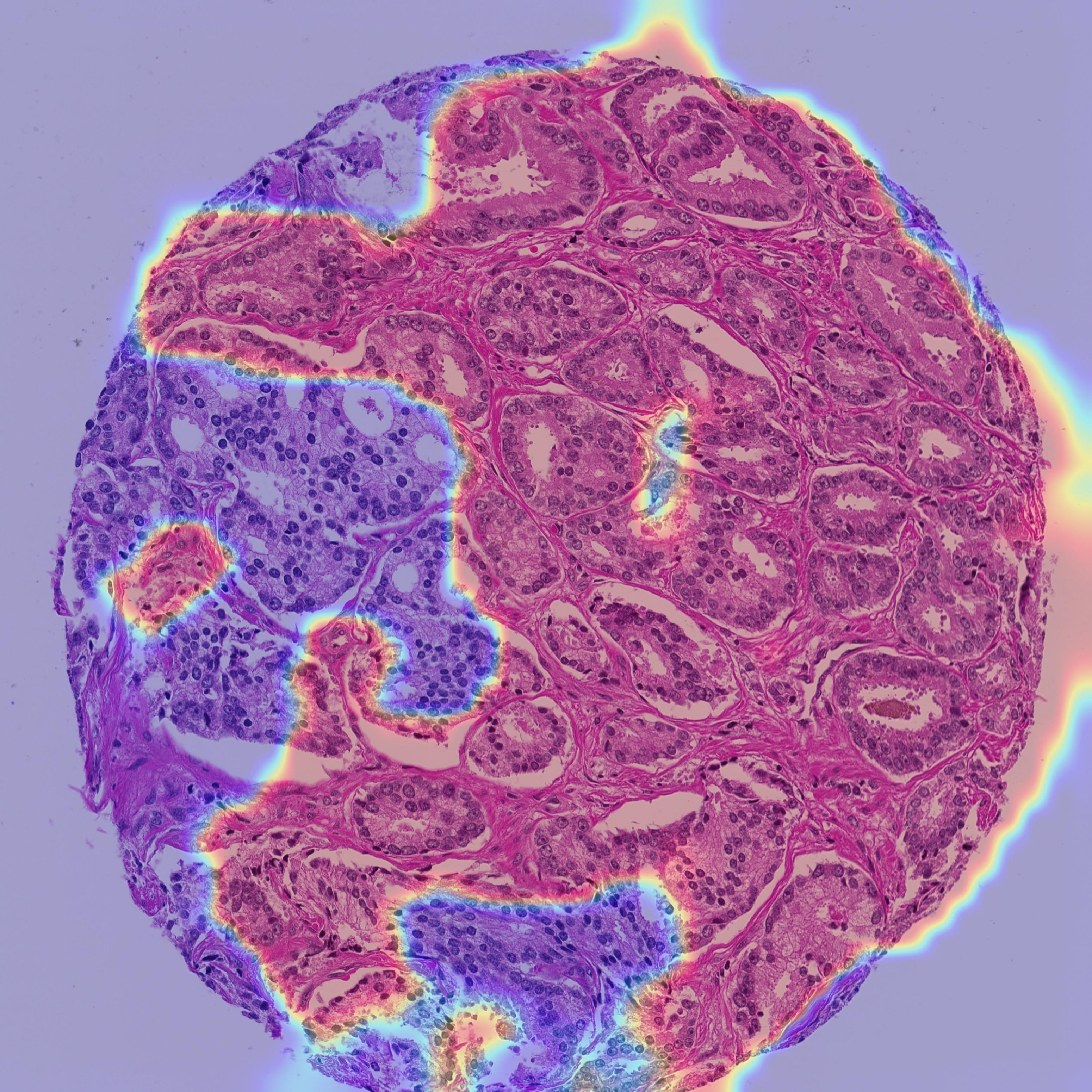}}
      \hspace*{\fill}
      \subfloat{\includegraphics[width=.19\linewidth]{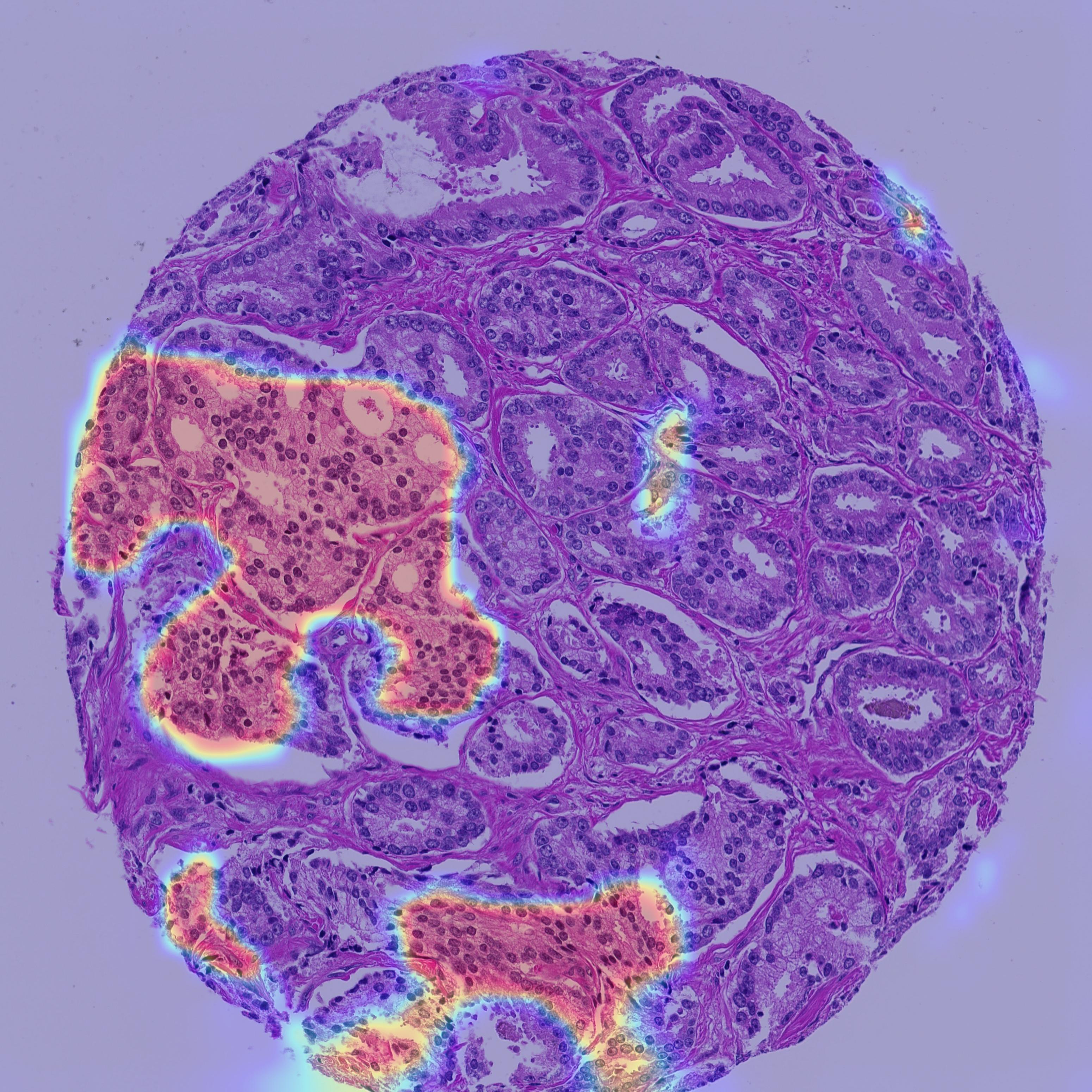}}
      \hspace*{\fill}
      \subfloat{\includegraphics[width=.19\linewidth]{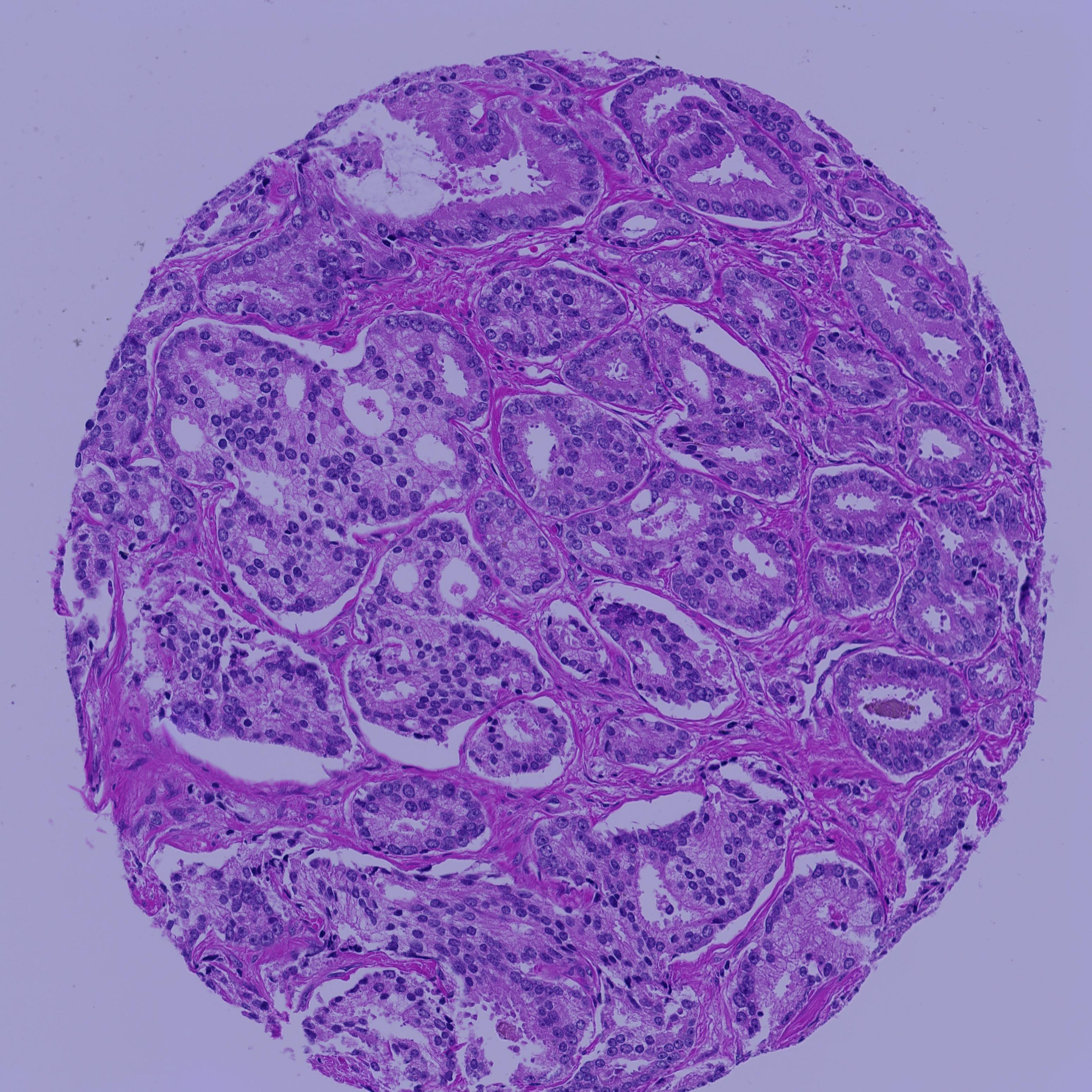}}
      \hspace*{\fill}
      \subfloat{\includegraphics[width=.19\linewidth]{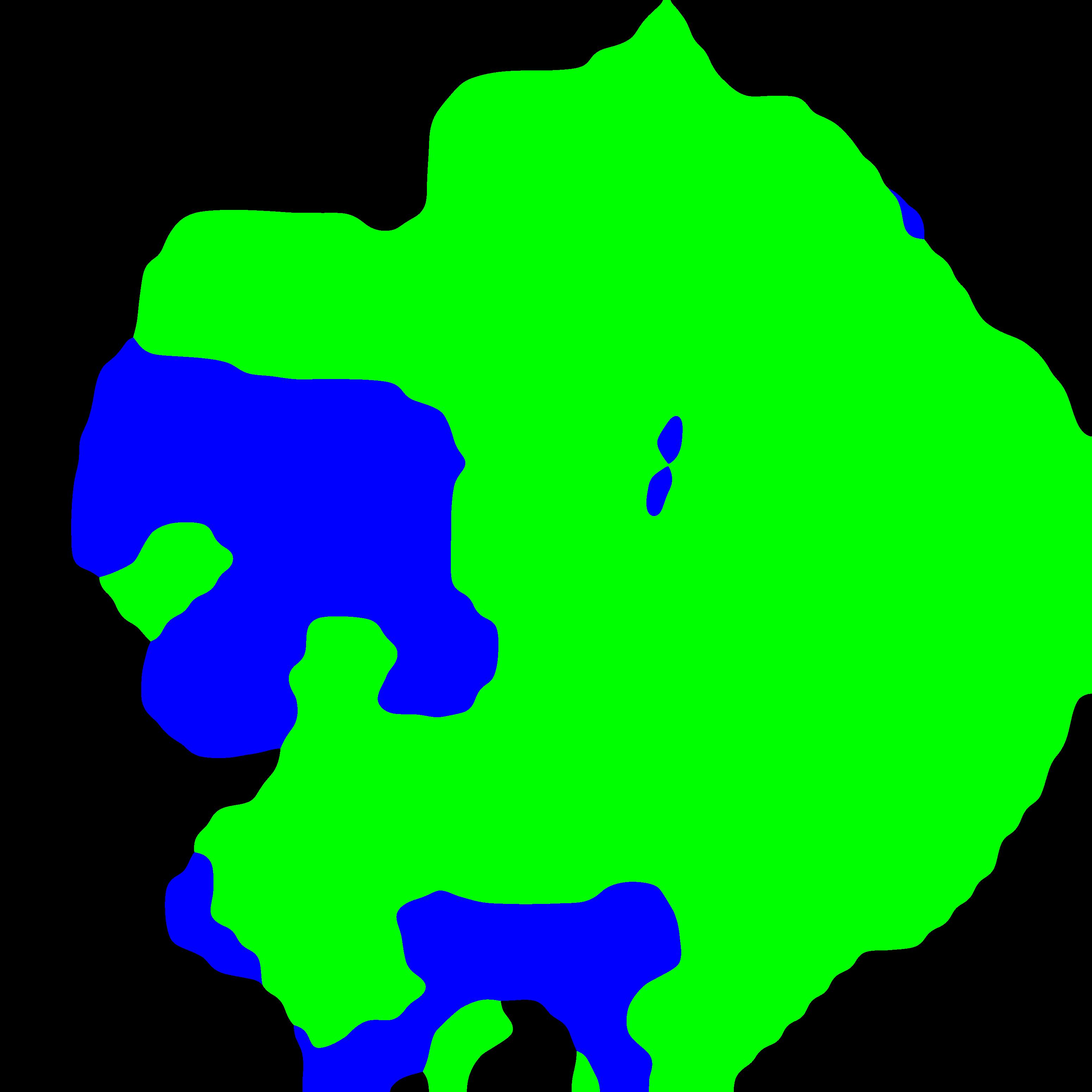}}
      \hspace*{\fill}    
    
      \subfloat{\includegraphics[width=.19\linewidth]{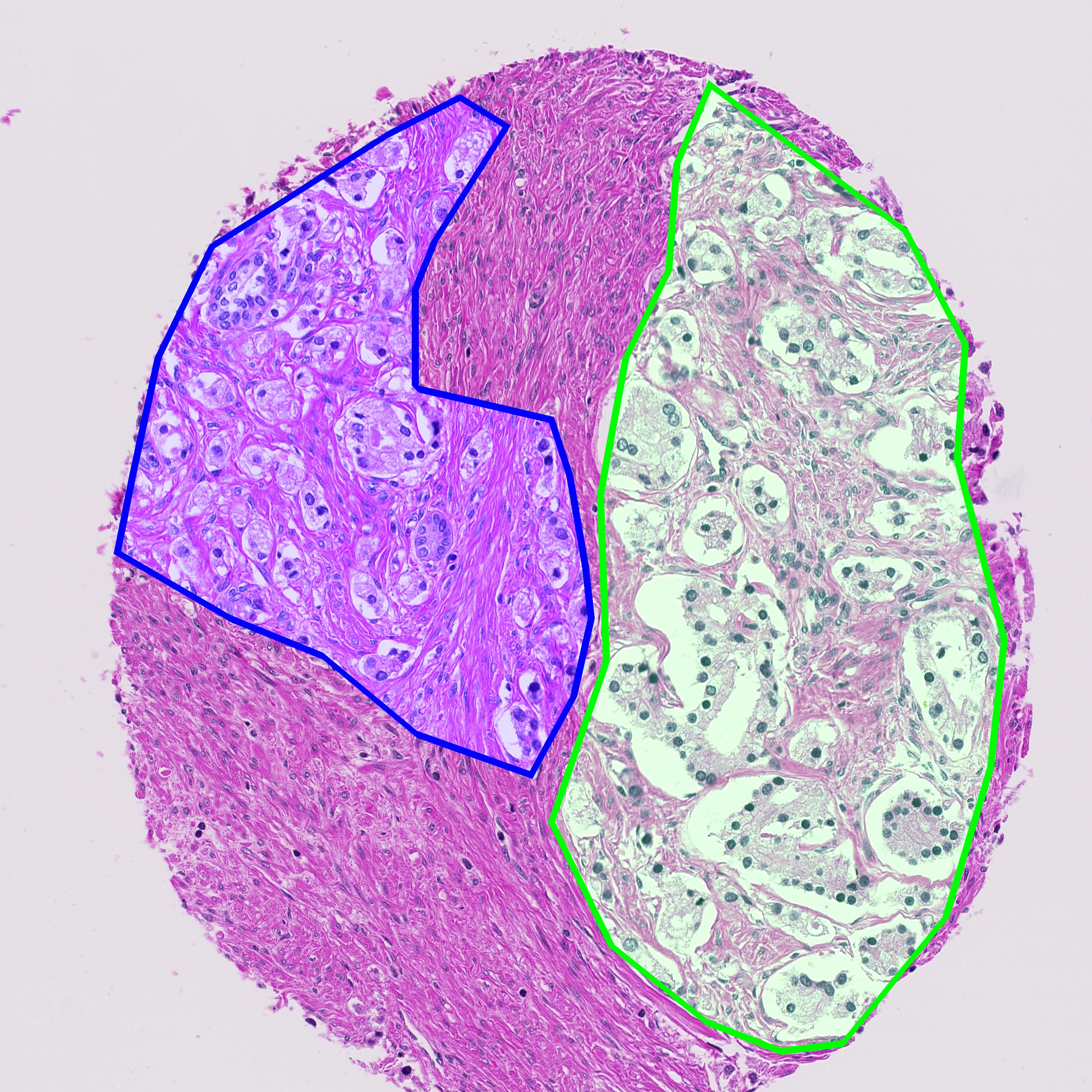}}
      \hspace*{\fill}
      \subfloat{\includegraphics[width=.19\linewidth]{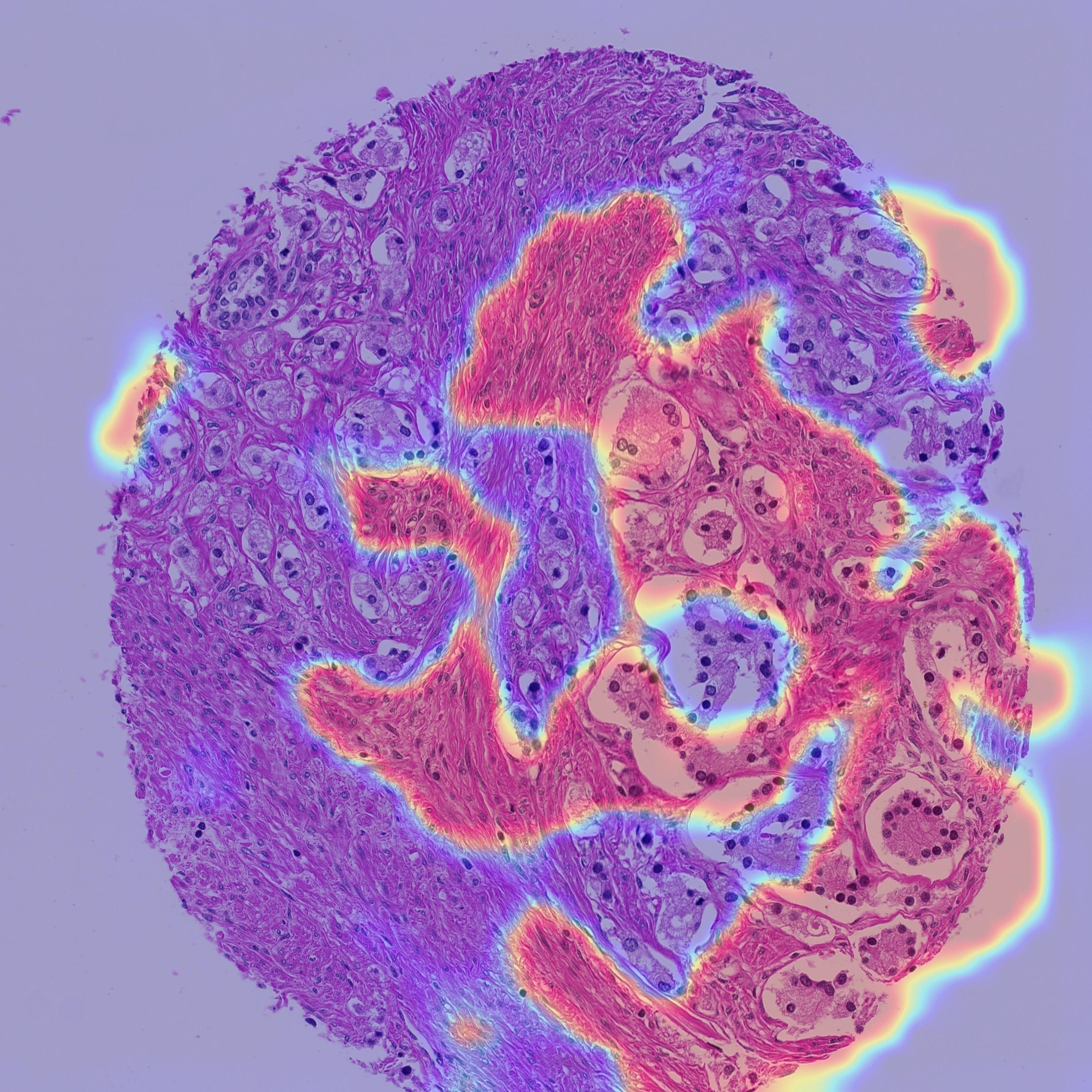}}
      \hspace*{\fill}
      \subfloat{\includegraphics[width=.19\linewidth]{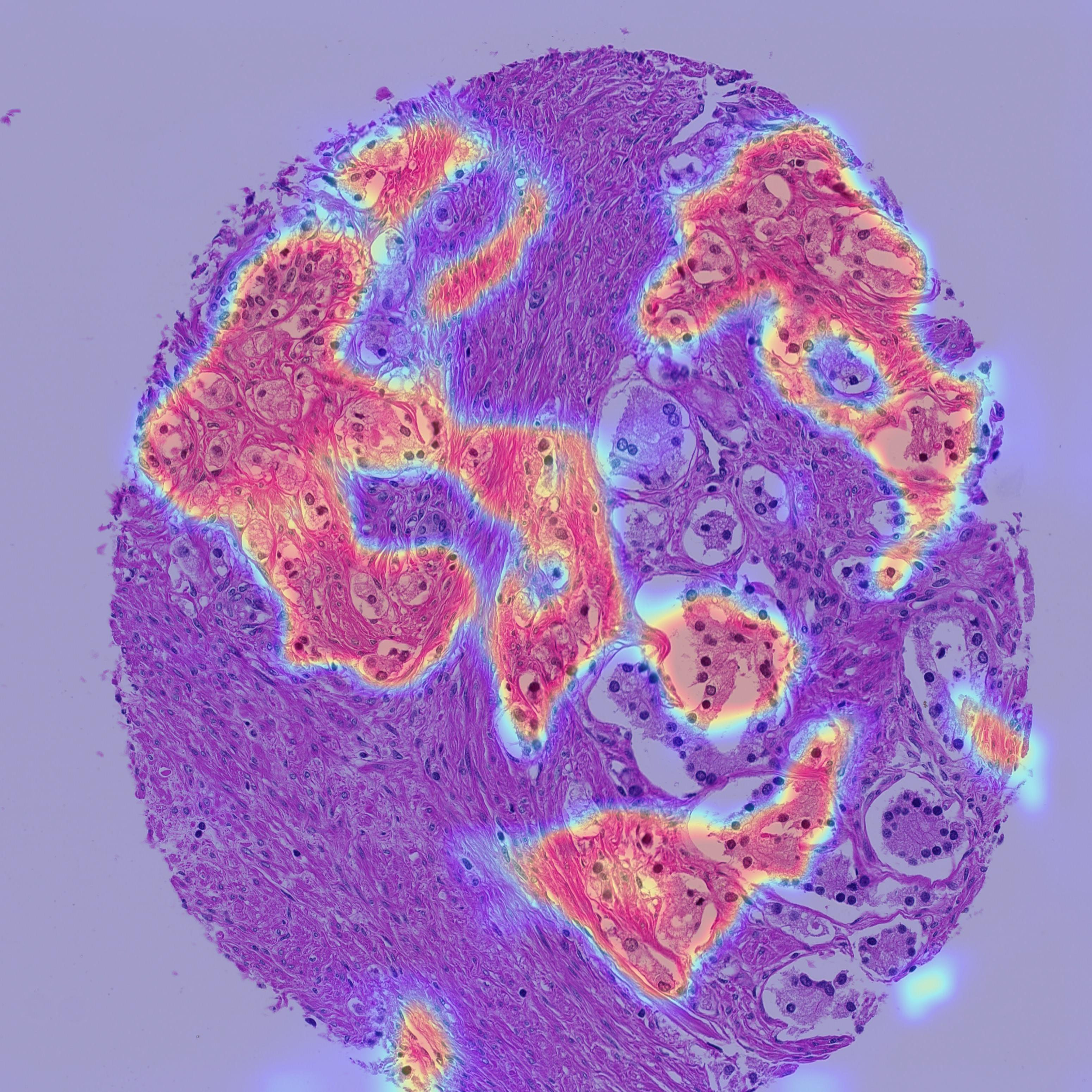}}
      \hspace*{\fill}
      \subfloat{\includegraphics[width=.19\linewidth]{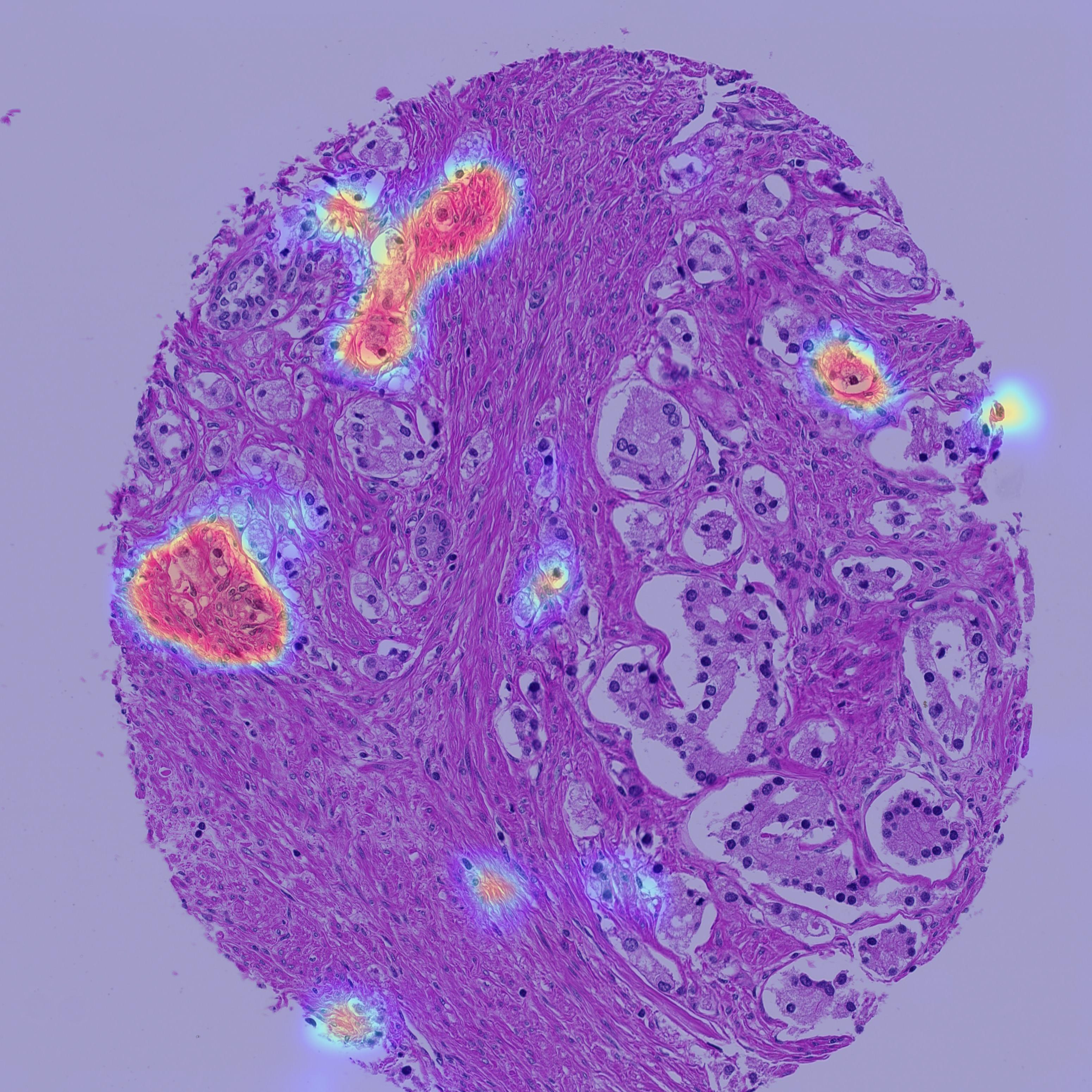}}
      \hspace*{\fill}
      \subfloat{\includegraphics[width=.19\linewidth]{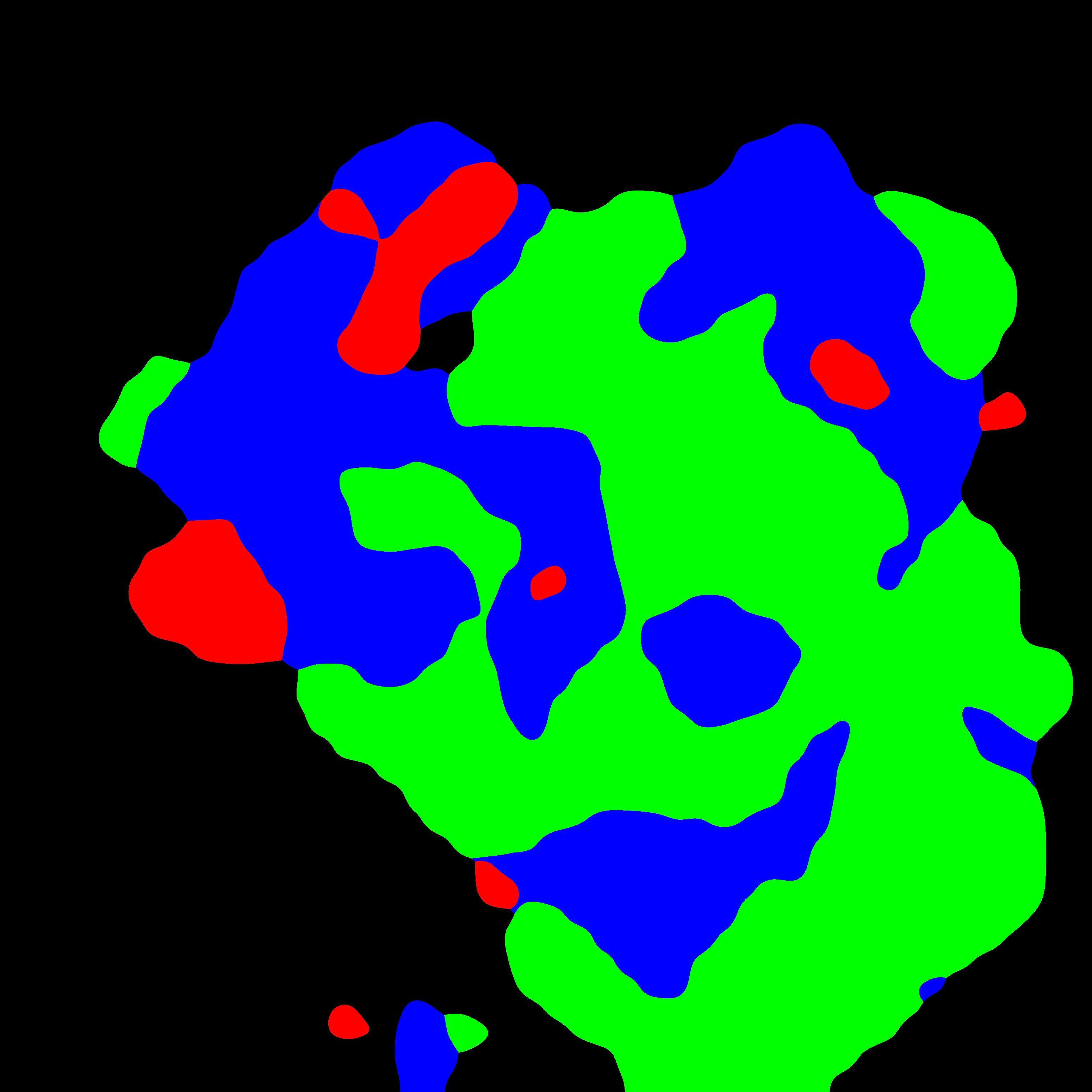}}
      \hspace*{\fill}

      \subfloat{\includegraphics[width=.19\linewidth]{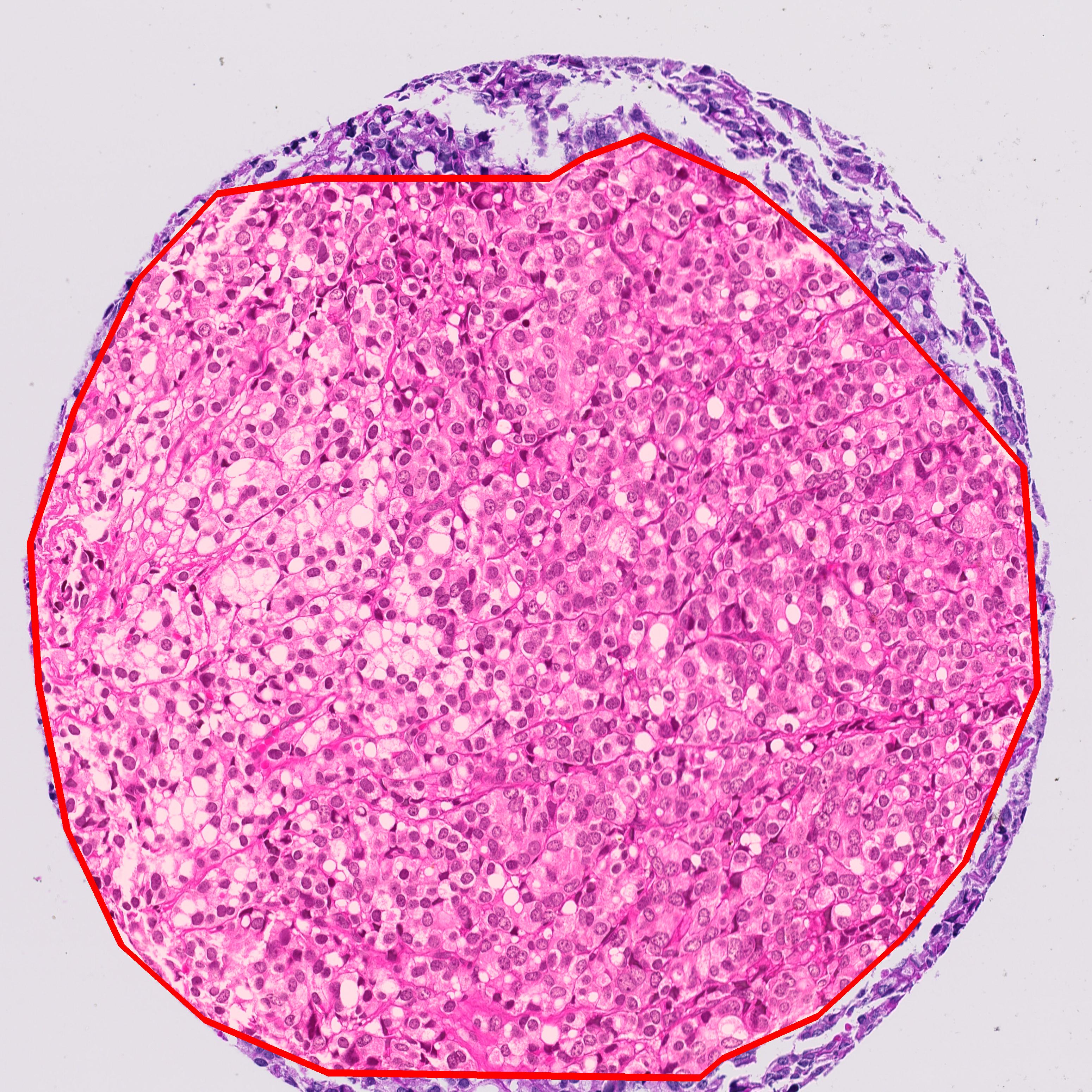}}
      \hspace*{\fill}
      \subfloat{\includegraphics[width=.19\linewidth]{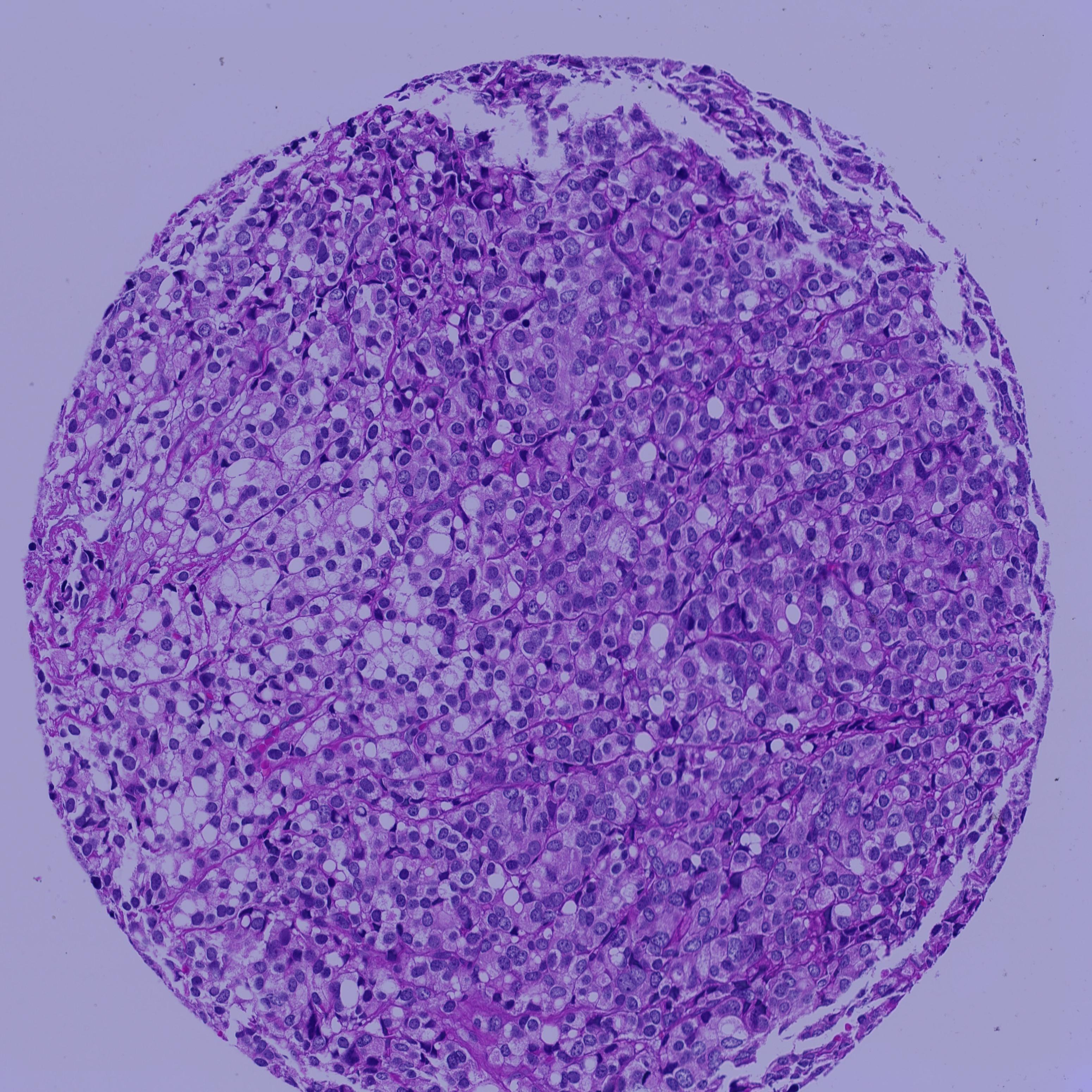}}
      \hspace*{\fill}
      \subfloat{\includegraphics[width=.19\linewidth]{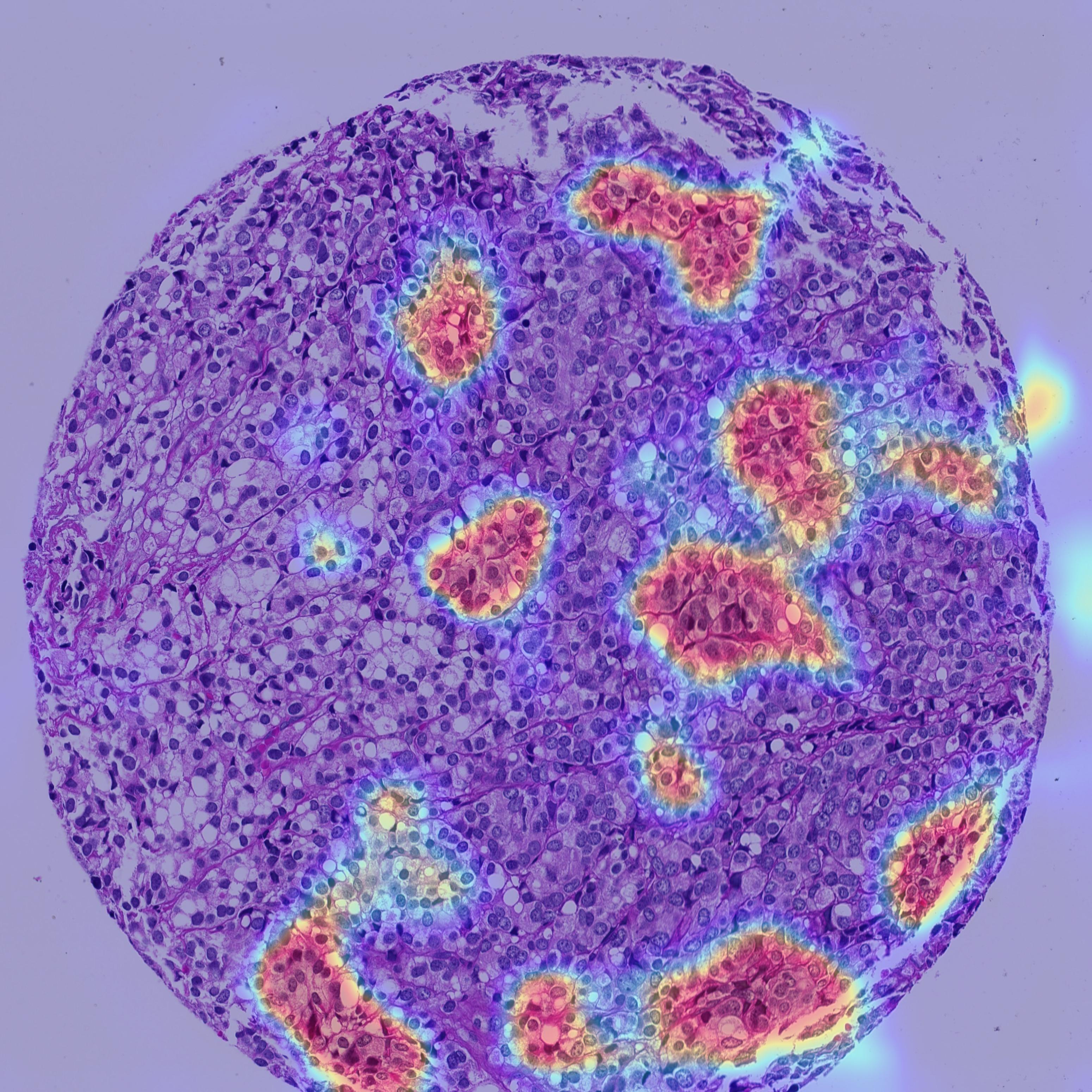}}
      \hspace*{\fill}
      \subfloat{\includegraphics[width=.19\linewidth]{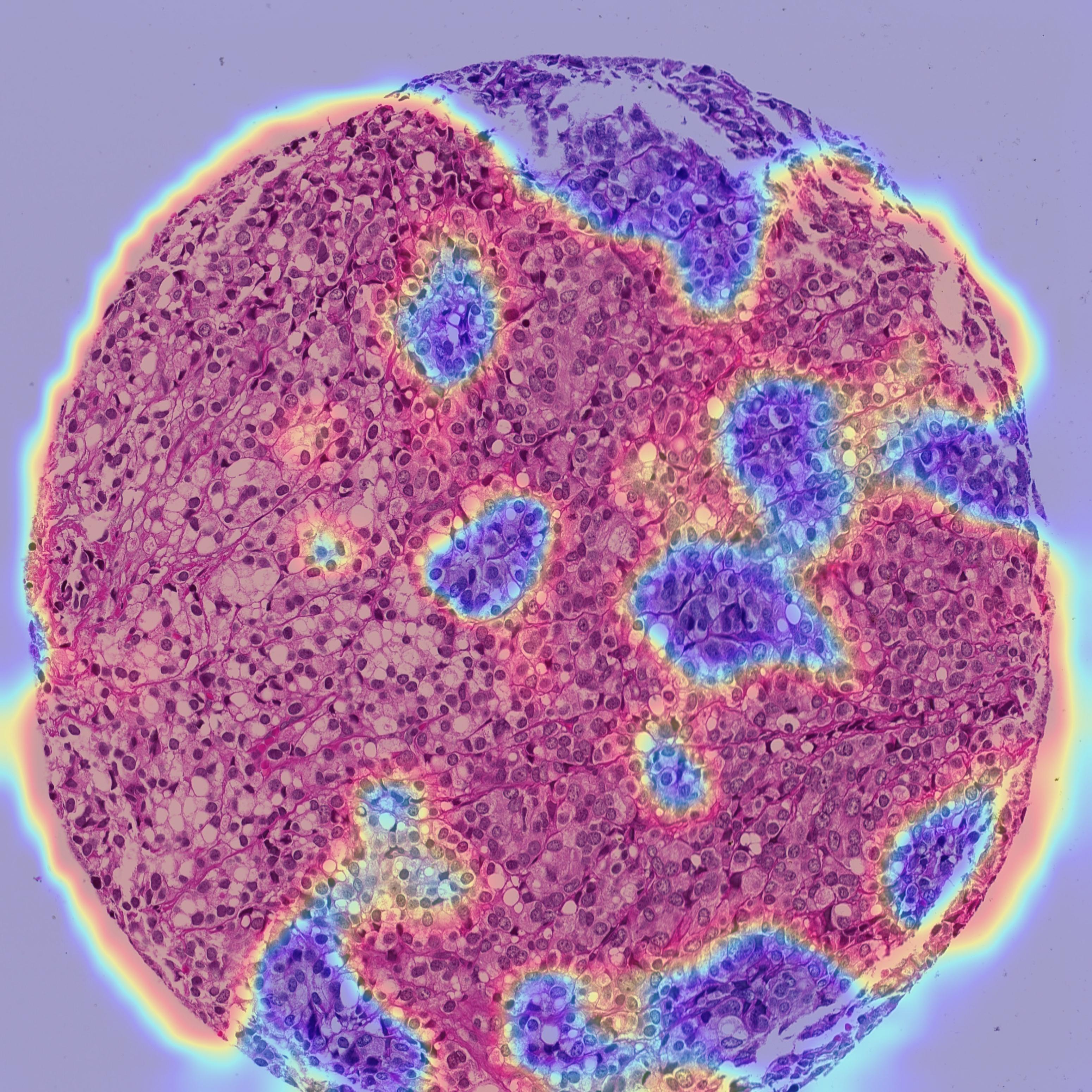}}
      \hspace*{\fill}
      \subfloat{\includegraphics[width=.19\linewidth]{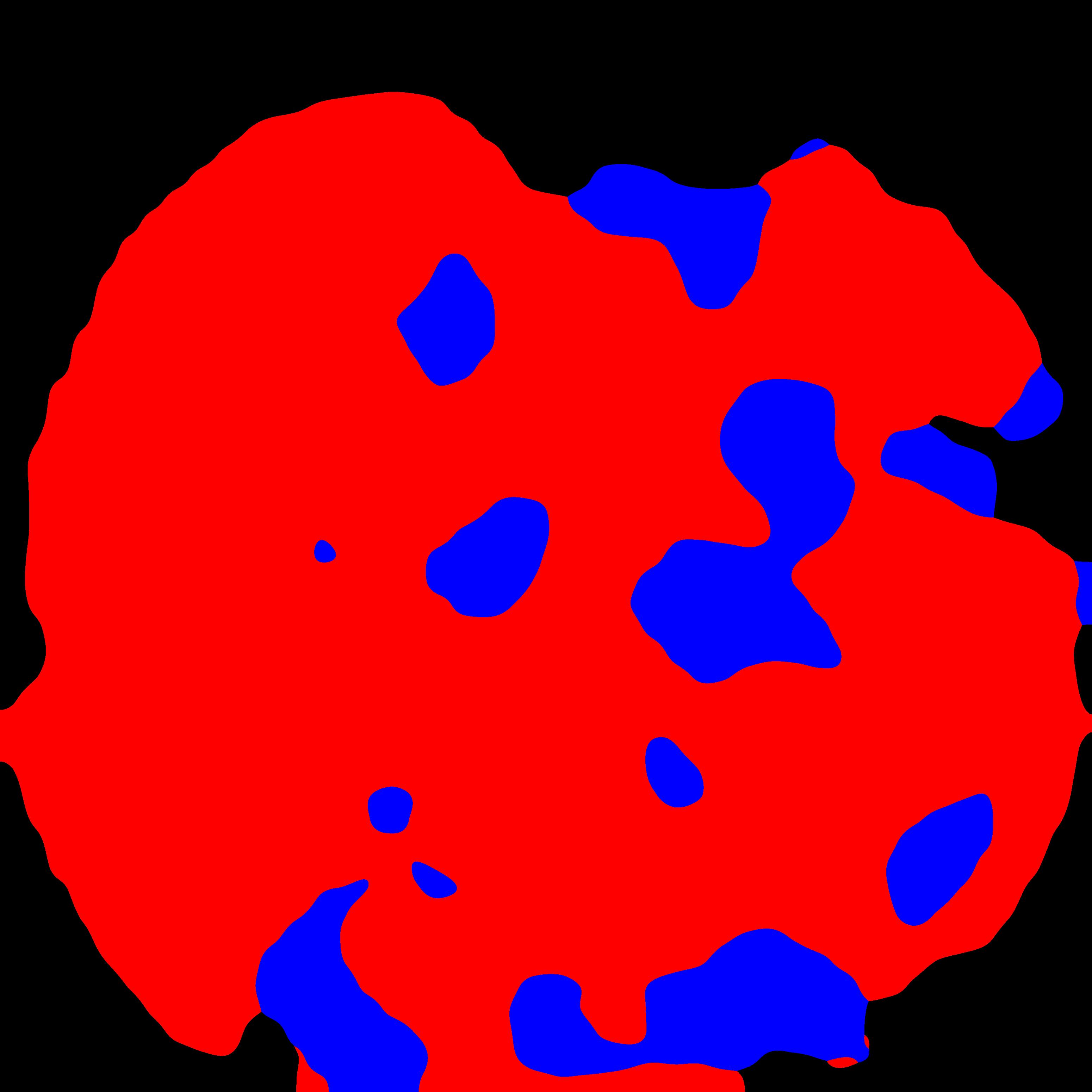}}
      \hspace*{\fill}    
    
      \subfloat{\includegraphics[width=.19\linewidth]{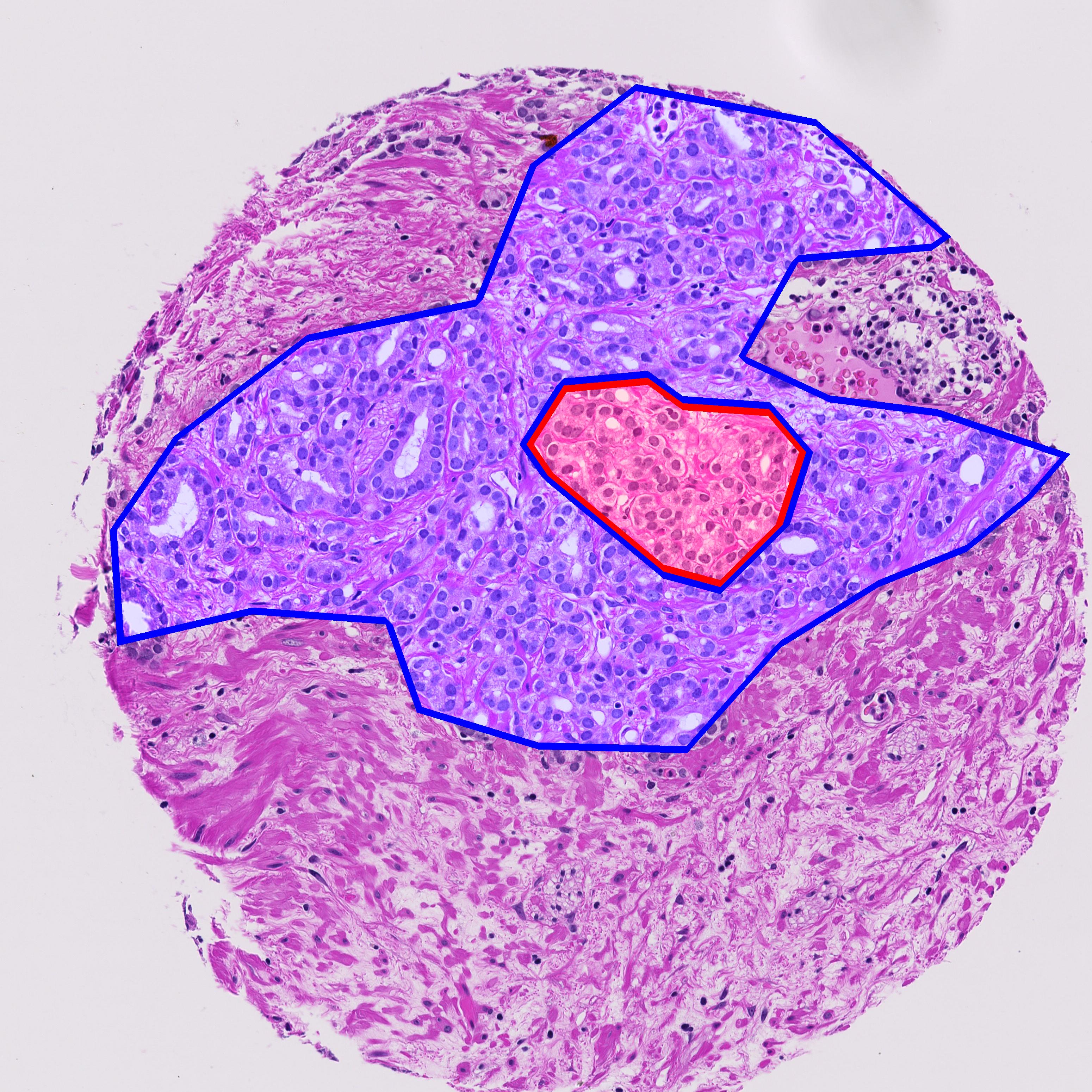}}
      \hspace*{\fill}
      \subfloat{\includegraphics[width=.19\linewidth]{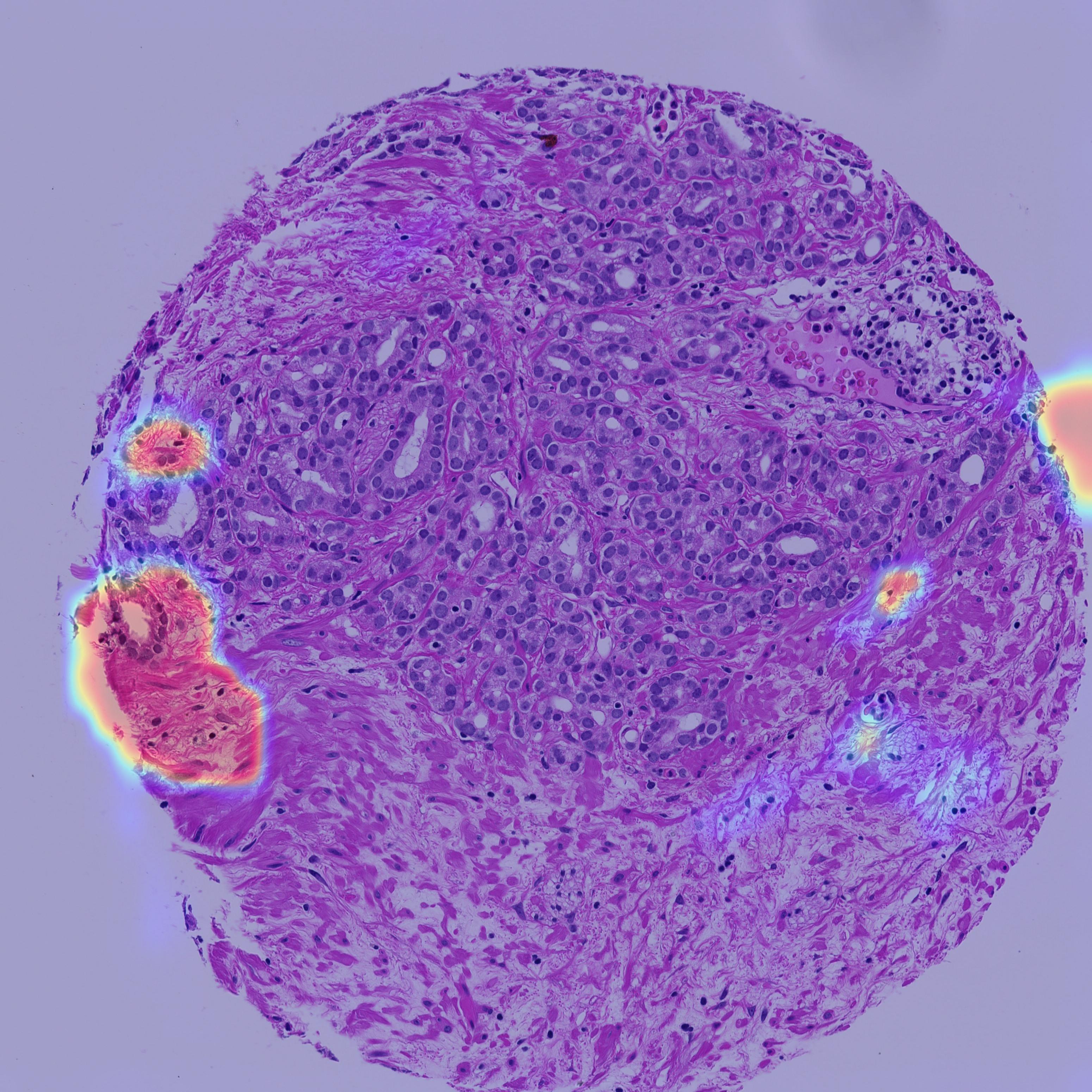}}
      \hspace*{\fill}
      \subfloat{\includegraphics[width=.19\linewidth]{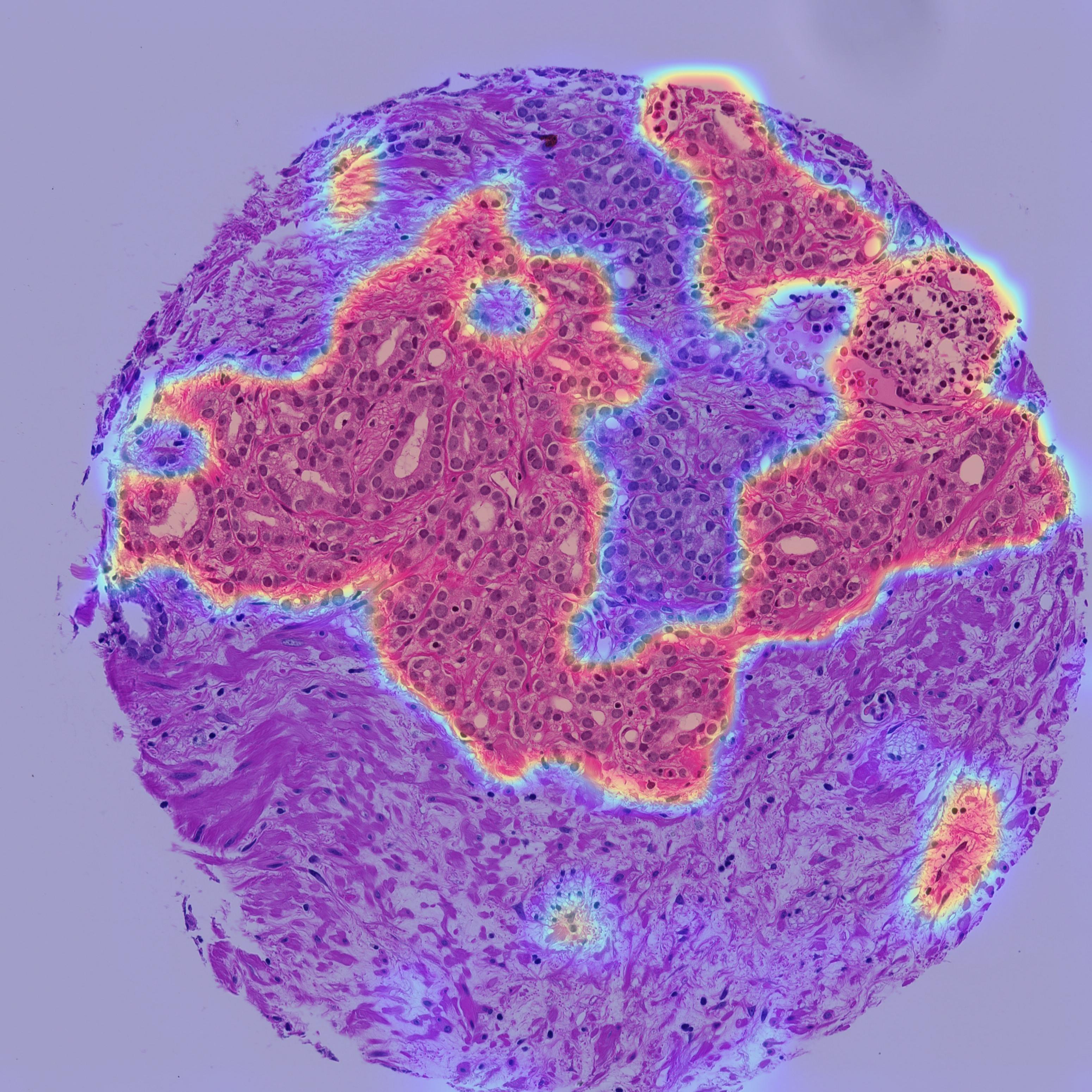}}
      \hspace*{\fill}
      \subfloat{\includegraphics[width=.19\linewidth]{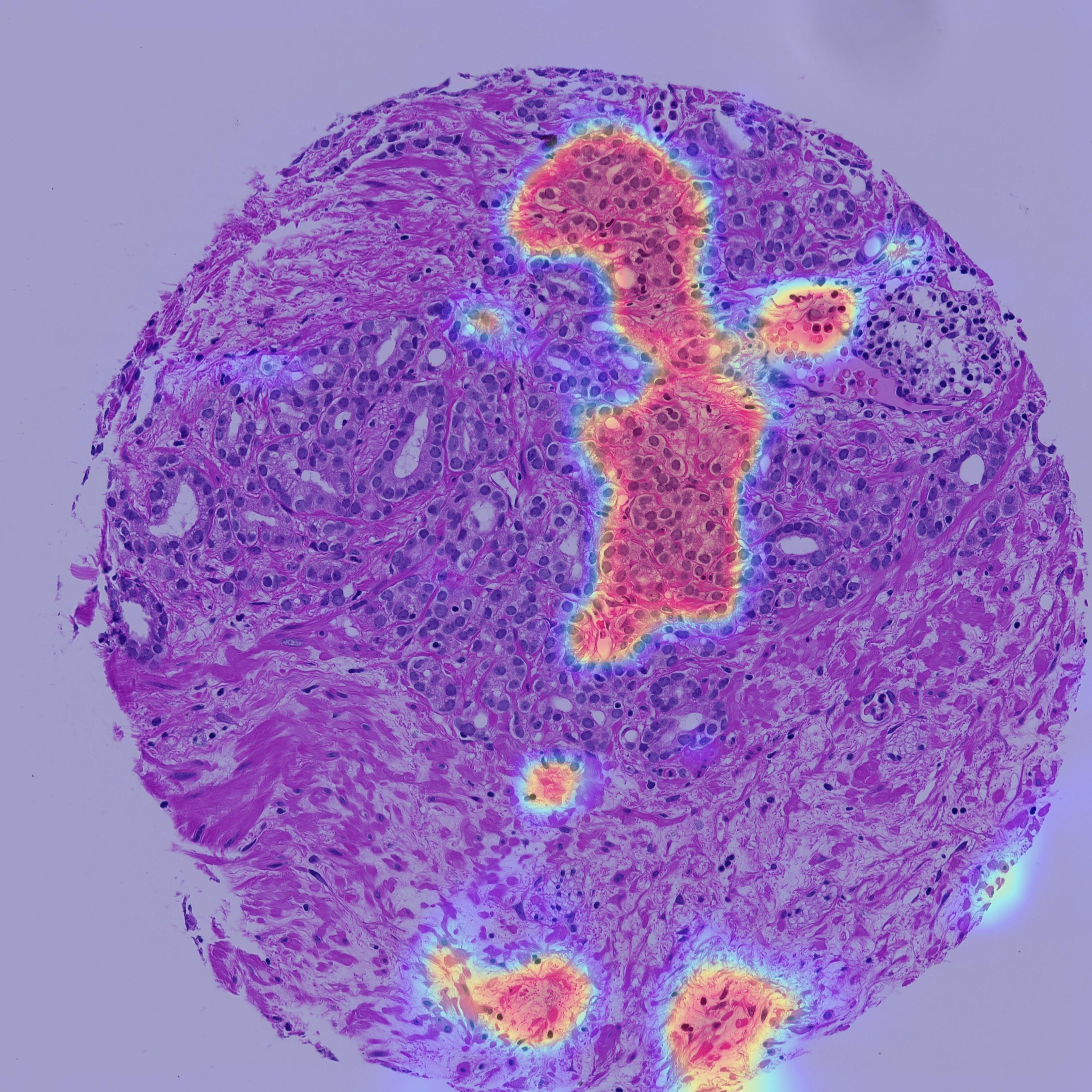}}
      \hspace*{\fill}
      \subfloat{\includegraphics[width=.19\linewidth]{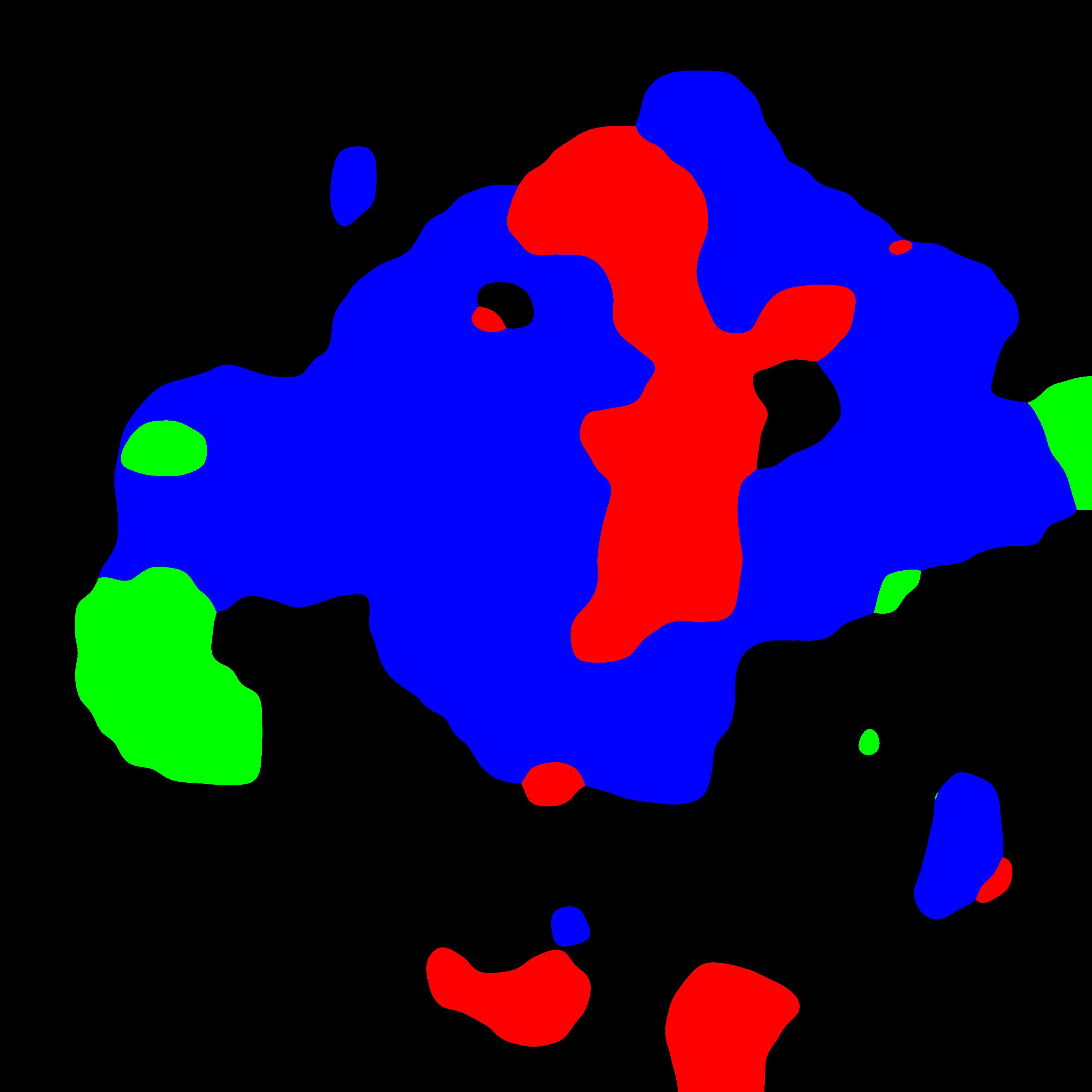}}
      \hspace*{\fill}    
    
      \renewcommand{\thesubfigure}{a}
      \subfloat[\label{a}]{\includegraphics[width=.19\linewidth]{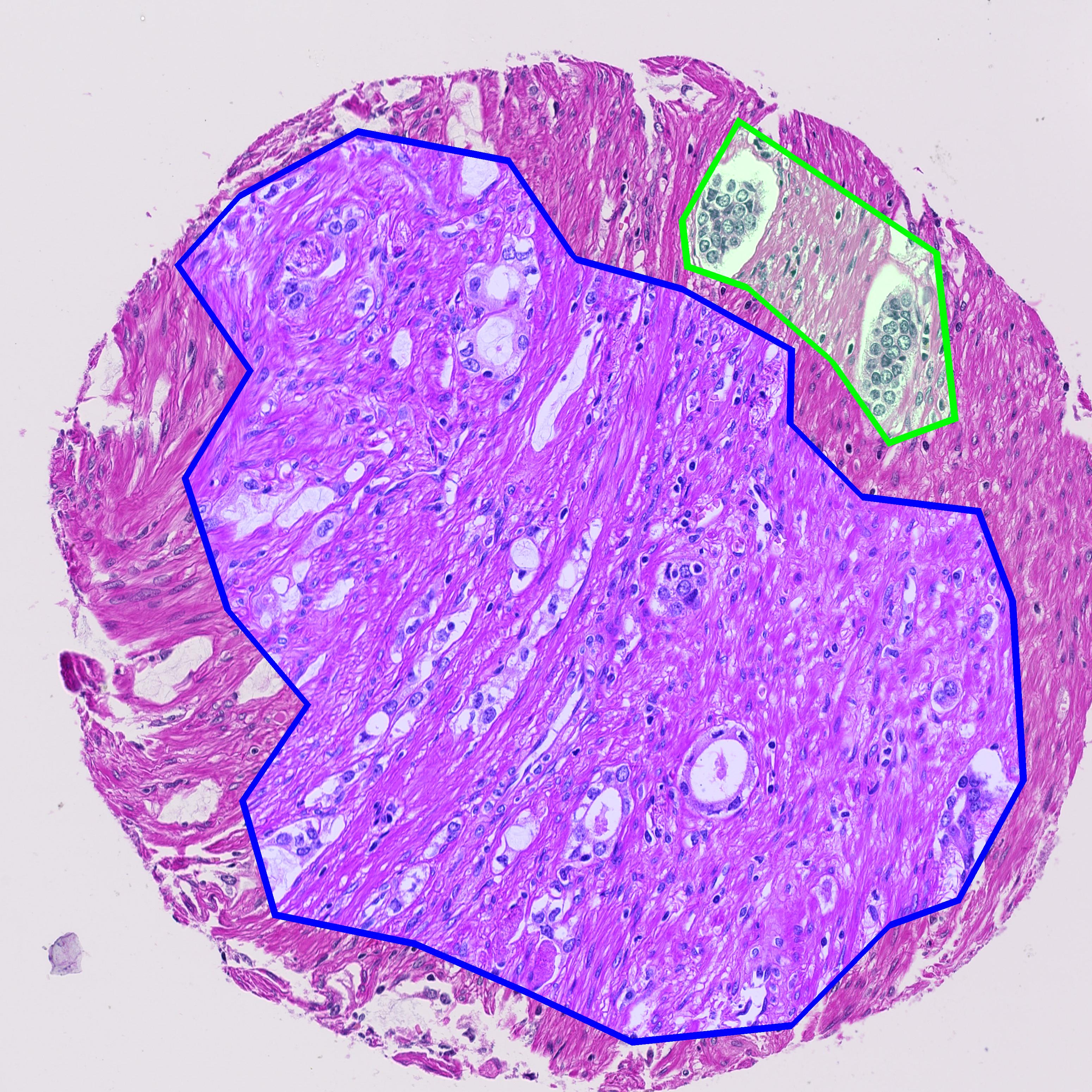}}
      \hspace*{\fill}
      \renewcommand{\thesubfigure}{b}
      \subfloat[\label{b}]{\includegraphics[width=.19\linewidth]{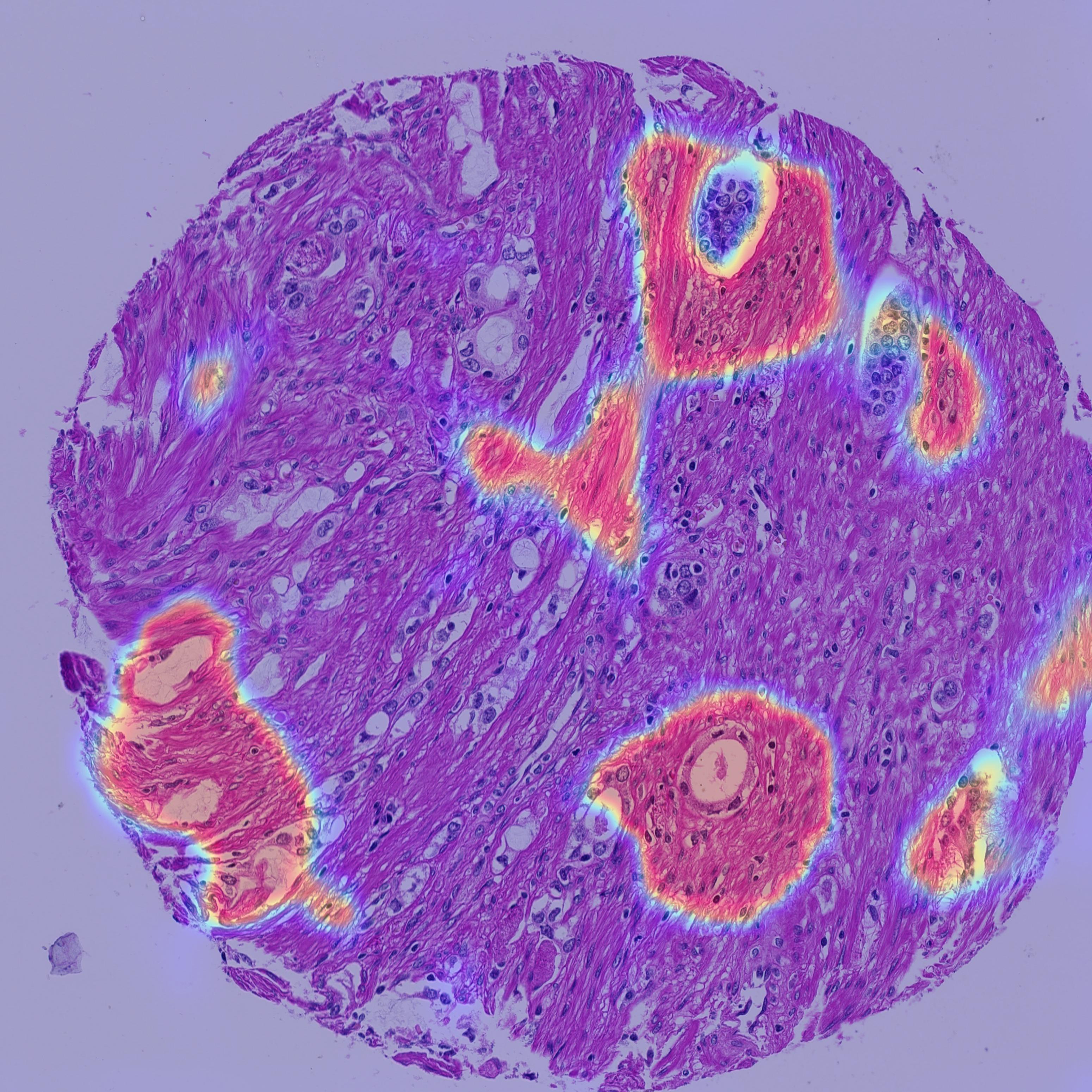}}
      \hspace*{\fill}
      \renewcommand{\thesubfigure}{c}
      \subfloat[\label{c}]{\includegraphics[width=.19\linewidth]{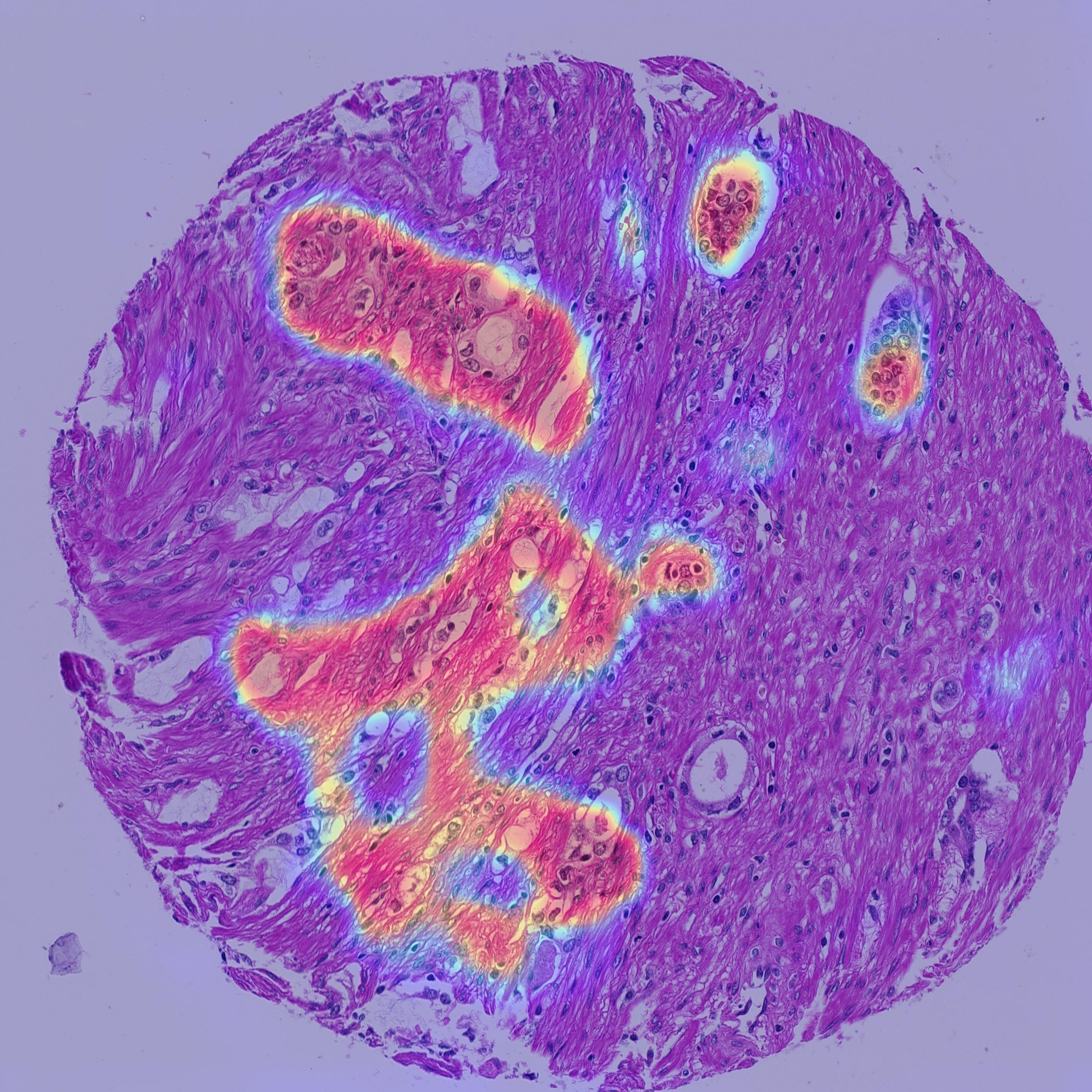}}
      \hspace*{\fill}
      \renewcommand{\thesubfigure}{d}
      \subfloat[\label{d}]{\includegraphics[width=.19\linewidth]{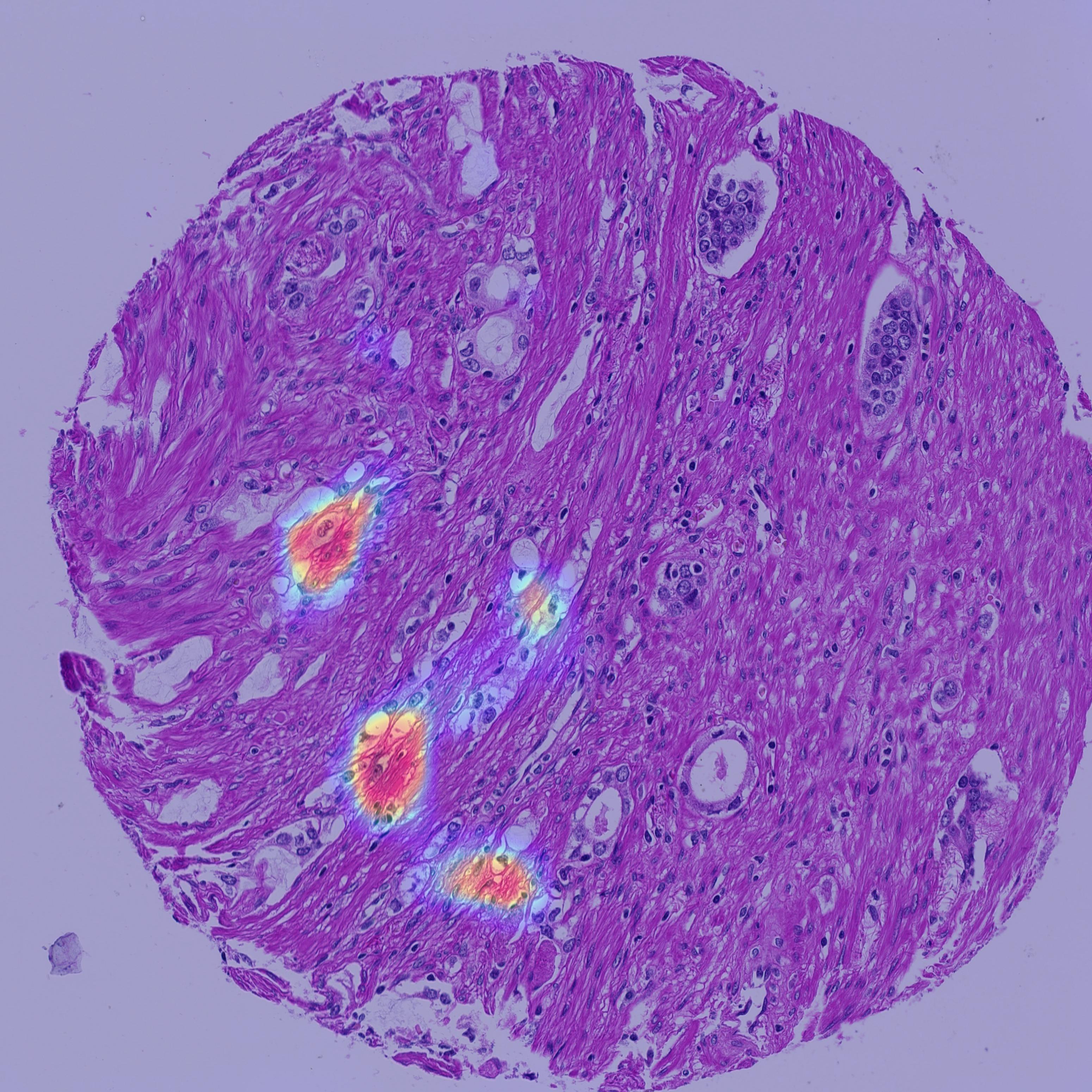}}
      \hspace*{\fill}
      \renewcommand{\thesubfigure}{e}
      \subfloat[\label{e}]{\includegraphics[width=.19\linewidth]{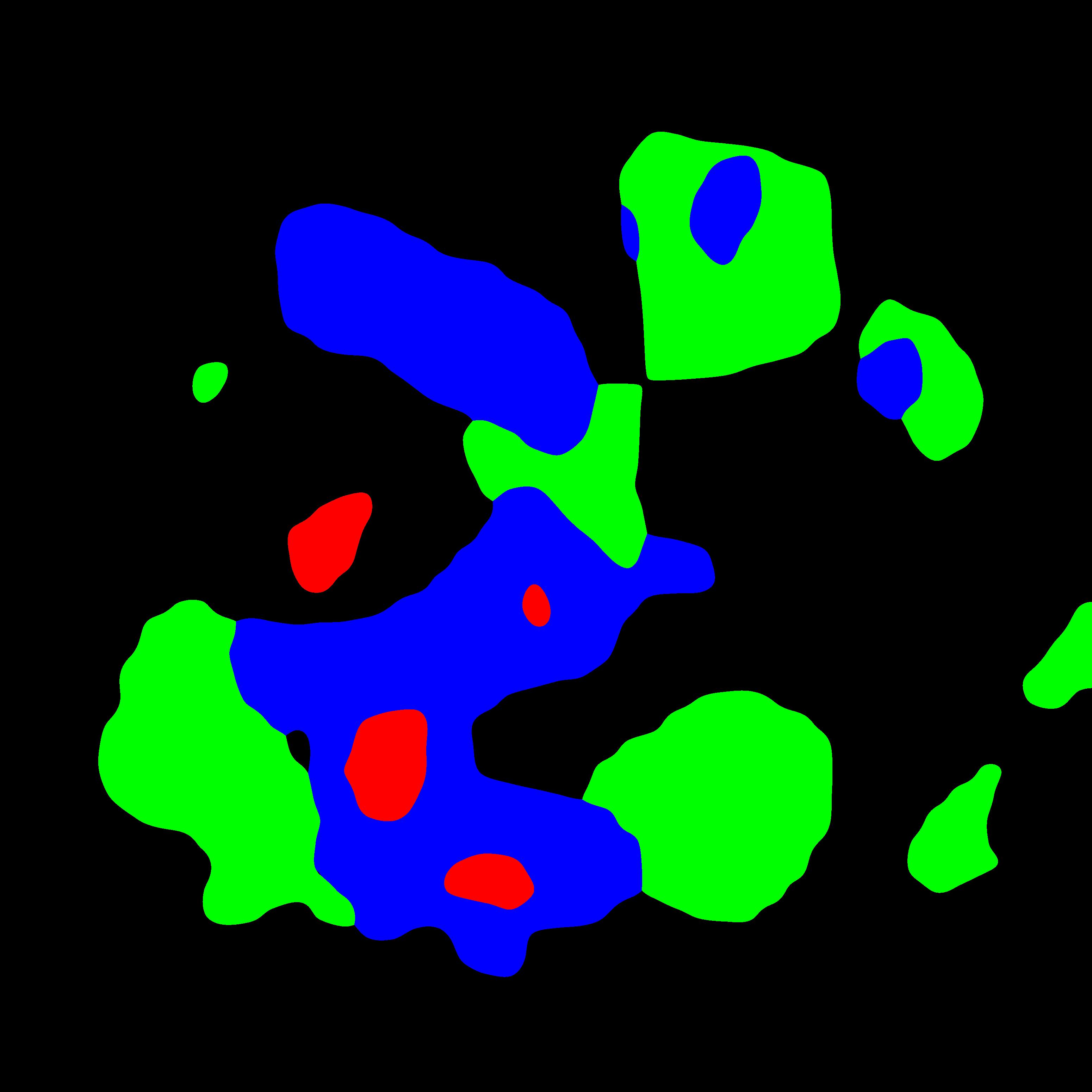}}
      \hspace*{\fill}
    
    \caption{Examples of the proposed weakly-supervised model, WeGleNet, segmentation performance in the test set. The reference annotations are obtained from Pathologist $1$. In green: Gleason grade $3$; blue: Gleason grade $4$ and red: Gleason grade $5$. (a): Reference; (b): Gleason grade $3$; (c): Gleason grade $4$; (d): Gleason grade $5$; (e): Semantic segmentation mask. The reference and predicted Gleason scores (Pathologist $1$ - Pathologist $2$ - Predicted), from top to bottom, are: ($6$ - $6$ - $6$); ($6$ - $7$ - $7$); ($7$ - $7$ - $8$); ($10$ - $10$ - $10$); ($8$ - $9$ - $8$) and ($7$ - $7$ - $7$).}
    \label{fig4}
\end{figure}

Once WeGleNet was optimized using the validation cohort, the best performing configuration, WeGleNet - $LSE_{r8}$ with HS regularization, was used to predict the segmentation maps from the images of the test cohort. Representative examples of the obtained results are presented in Figure \ref{fig4}. This figure is organized as follows: each row is a different core and each column represents the ground truth of the Pathologist $1$, and the predicted heatmaps for GG3, GG4 and GG5 classes, respectively. Finally, the last column presents the discrete-valued semantic segmentation maps, assigning to each pixel the class with the highest probability. In this figure, green, blue and red color indicate GG3, GG4 and GG5 patterns, respectively.

% --------------------------------
%% Figure with graphics examples

Then, we carried out experiments to compare our proposed weakly-supervised model with respect to the state-of-the-art supervised methods. UNet model was trained using Nadam as optimizer, with a learning rate of $1\cdot10^{-4}$ during $60$ epochs. In each iteration, a mini-batch of $16$ images was used to update the models weights. Regarding the VGG19Sup model, a learning rate of $1\cdot10^{-3}$ with SGD as optimizer was used. Training was performed in mini-batches of 64 images during $120$ epochs. For both models, early stopping was applied to keep the best-performing model in the validation cohort (in terms of the obtained loss). From the trained models, the segmentation maps of the images in the test cohort were predicted. The figures of merit obtained by our proposed WeGleNet - $LSE_{r8}$ and the supervised models are presented in Table \ref{tab:supVSweak}, using as reference the annotations carried out by the pathologist $1$ (the same pathologist that annotated the training and validation images). In order to perform a detailed comparison, accuracy (ACC), class-level f1-score (F1), average intersection over union (mIoU) and quadratic Cohen's kappa ($\kappa$) were obtained as detailed in Section \ref{ssec:exp1}.

% --------------------------------
%% Table with figures of merit

\begin{table}[htb]

    \centering
    \caption{Results of the Gleason grades semantic segmentation using the proposed weakly-supervised model, WeGleNet, and two supervised approaches, SupVGG19 and UNet. The metrics presented are the accuracy (ACC), the F1-Score (F1), computed per class and its average, the mean intersection over union ($mIoU$) and the Cohen's quadratic kappa ($\kappa$).}
    \label{tab:supVSweak}
    \resizebox{\linewidth}{!}{
    \begin{tabular}{|l|c|ccccc|c|c|}
    \hline
    \multicolumn{1}{|c|}{\textbf{Experiment}} & \textbf{ACC} & \multicolumn{5}{|c|}{\textbf{F1}} & \textbf{mIoU} & \textbf{$\kappa$}\\
     & & NC & GG3 & GG4 & GG5 & Avg. & &  \\
    \hline
    
    WeGleNet - $LSE_{r8}$ & $0.6859$ & $0.8155$ & $0.5883$ & $0.5622$ & $\mathbf{0.3531}$ & $\mathbf{0.5798}$ & $\mathbf{0.4368}$ & $0.6105$\\
    \hline
    SupVGG19 & $0.5426$ & $0.6747$ & $0.5195$ & $0.4959$ & $0.1551$ & $0.4613$ & $0.3497$ & $0.2630$\\
    \hline
    UNet & $\mathbf{0.6968}$ & $\mathbf{0.8383}$ & $\mathbf{0.5932}$ & $\mathbf{0.5737}$ & $0.2419$ & $0.5618$ & $0.4178$ & $\mathbf{0.6387}$\\
    
    \hline
    \end{tabular}
    }
\end{table}

WeGleNet - $LSE_{r8}$ model reached a $\kappa$ value of $0.6105$, a $mIoU$ f $0.4368$ and an average F1 of $0.5798$ in the semantic segmentation of Gleason grades in the test cohort. Our proposed model outperformed the supervised SupVGG19 model segmentation ($\kappa = 0.2630$, $mIoU = 0.3497$ and $F1 = 0.4613$), and it performs similarly to the UNet model ($\kappa = 0.6387$, $mIoU = 0.4178$ and $F1 = 0.5618$). Although the UNet model reached better results in the non-cancerous class ($F1 = 0.8383$), WeGleNet - $LSE_{r8}$ differentiated better the Gleason grades, reaching an F1 of $0.3531$ for the GG5 class, a challenging task due to the low prevalence of these patterns. Thus, our proposed WeGleNet model performed at a level equivalent to supervised methods in the segmentation of Gleason grades, without requiring pixel-level annotations.

% --------------------------------
%% Comparison with SoA

\subsection{State-of-the-Art Comparison}
\label{ssec:exp4}

\begin{figure}[H]
\captionsetup[subfloat]{farskip=1pt,captionskip=0.8pt}
    \begin{center}
    
      \subfloat[\label{fig:cm1a}]{\includegraphics[width=.49\linewidth]{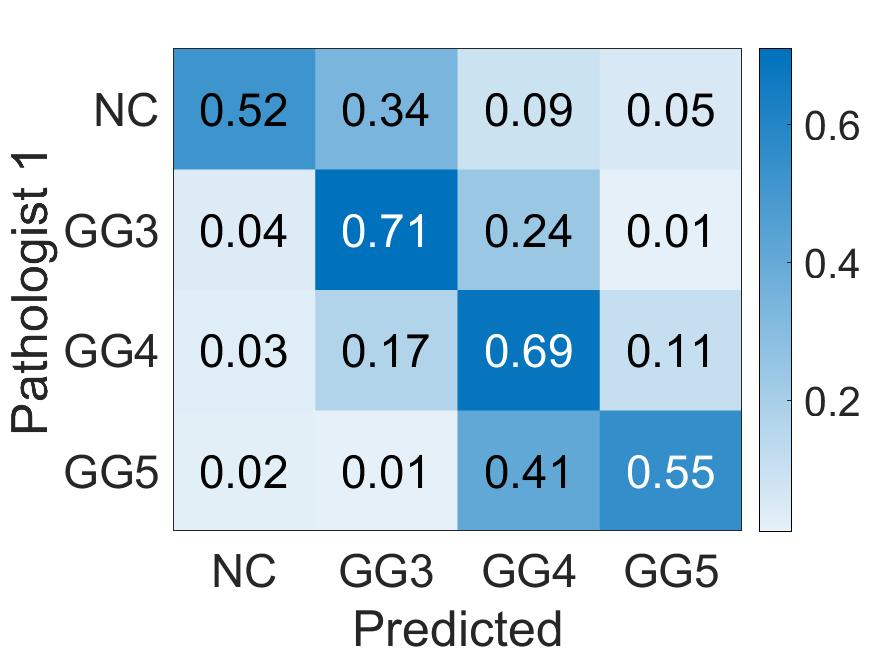}}
      \hspace*{\fill}
      \subfloat[\label{fig:cma}]{\includegraphics[width=.49\linewidth]{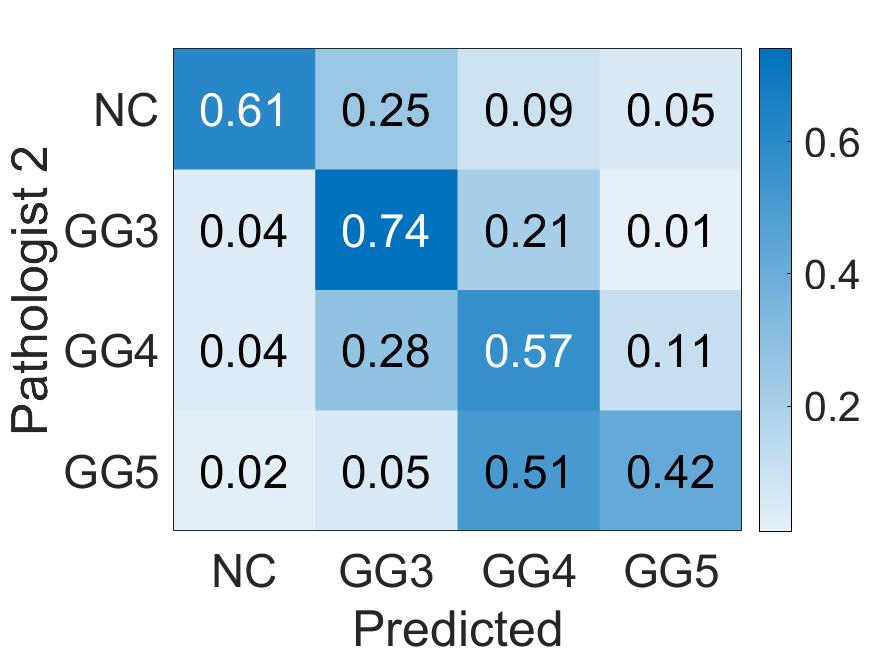}}
      \hspace*{\fill}

    \caption{Confusion Matrix of the patch-level Gleason grades prediction done by WeGleNet - $LSE_{r8}$ network in the test subset. The reference labels in each matrix are obtained from: (a) pathologist $1$, and (b) pathologist $2$. GG: Gleason grade; NC: non cancerous.}
    \label{fig:cm}
    \end{center}
\end{figure}

Finally, predictions were obtained at patch-level (which extraction is specified in Section \ref{ssec:exp1}) to compare WeGleNet against previous works in the used dataset. In the test cohort, patch-level classifications were obtained by majority voting of pixel-level predictions. Only fully non-cancerous patches were predicted as benign. The Cohen's quadratic kappa ($\kappa$) was obtained using the annotations of both pathologists. The figures of merit are presented in Table \ref{tab:soaVSweak} and confusion matrices are presented in Figure \ref{fig:cm}.

\begin{figure}[H]
\captionsetup[subfloat]{farskip=1pt,captionskip=0.8pt}
    \begin{center}
    
      \subfloat[\label{fig:cm2a}]{\includegraphics[width=.49\linewidth]{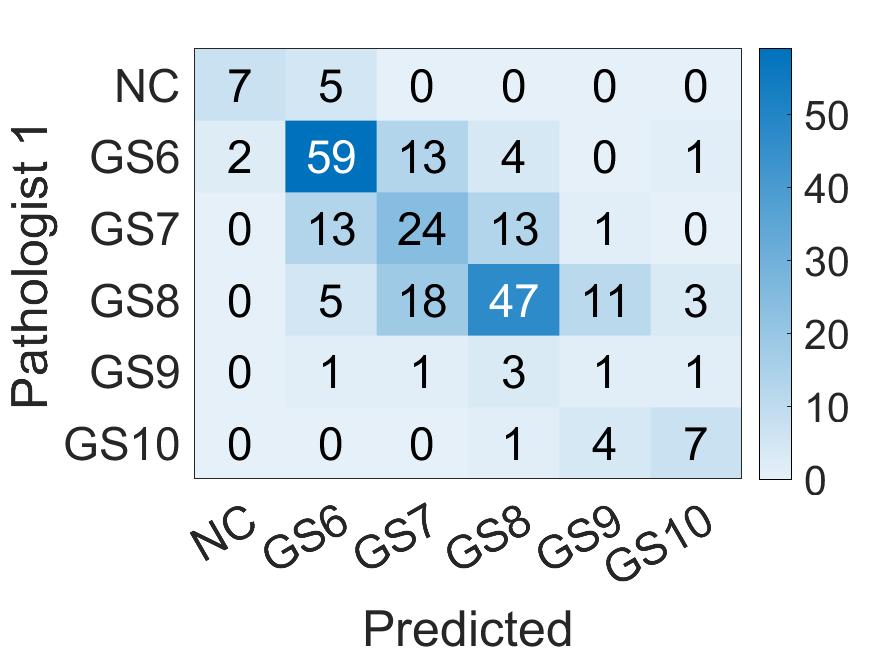}}
      \hspace*{\fill}
      \subfloat[\label{fig:cm2b}]{\includegraphics[width=.49\linewidth]{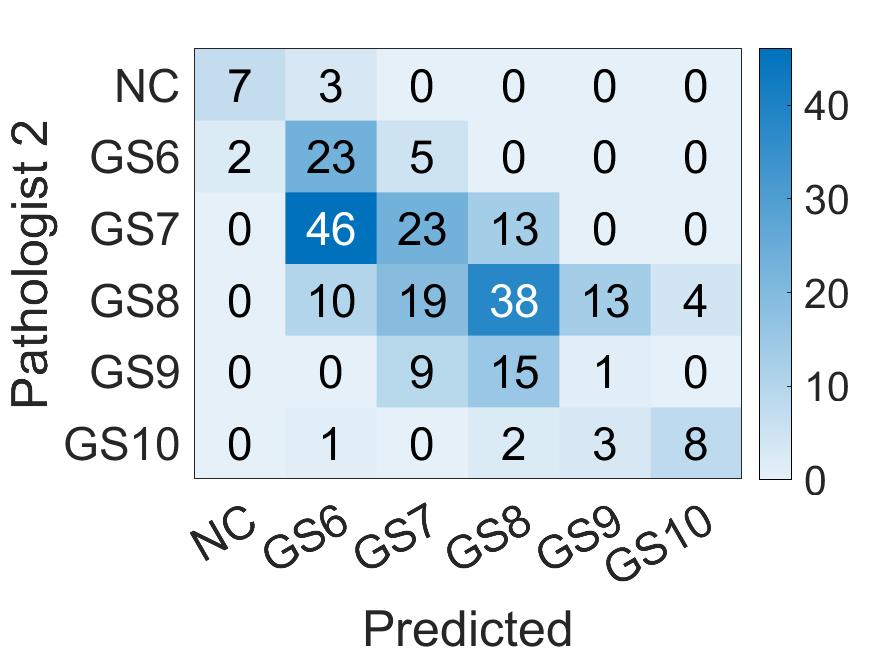}}
      \hspace*{\fill}

    \caption{Confusion Matrix of the global-level Gleason scores prediction done by WeGleNet network in the test subset. The reference labels in each matrix are obtained from: (a) pathologist $1$, and (b) pathologist $2$.}
    \label{fig:cm2}
    \end{center}
\end{figure}

Then, the global Gleason scoring of the cores was performed as described in Section \ref{ssec:meth4}. The parameters $c=0.03$ and $d=0.70$ were empirically fixed using the validation set. The $\kappa$ and confusion matrices were obtained using as reference both pathologists, and the results are reported in Table \ref{tab:soaVSweak} and Figure \ref{fig:cm2}, respectively. Moreover, the obtained Gleason Score and references of representative cores are indicated in Figure \ref{fig4}.

In order to compare the obtained figures of merit with previous literature, the reported results for the patch-level grading and global scoring obtained using fully-supervised models with pixel-level annotations by Arvaniti et al. \cite{Arvaniti2018AutomatedLearning} are indicated in Table \ref{tab:soaVSweak}. Also, the results obtained in this test set by Bulten et al. \cite{Bulten2020AutomatedStudy} using semi-supervised models trained in a large set of biopsies (see Section \ref{ssec:rw3} for a more detailed description) are pointed out in that table.

\begin{table}[htb]
\centering
\caption{Results of the patch-level Gleason grading and core-level scoring of the proposed model and comparison with previous literature. The metric presented is the Cohen's quadratic kappa ($\kappa$).}
\label{tab:soaVSweak}
\begin{tabular}{|l|c|c|}
\hline
\multicolumn{1}{|c|}{{\textbf{Approach}}} & \multicolumn{2}{c|}{\textbf{$\kappa$}}                                                           \\ 
\cline{2-3} 
\multicolumn{1}{|c|}{}                                   & \multicolumn{1}{l|}{\textbf{Pathologist 1}} & \multicolumn{1}{l|}{\textbf{Pathologist 2}} \\
\hline
\hline
\multicolumn{3}{c}{Patch-Level Grading} \\
\hline
\hline
WeGleNet                                         & $0.59$                                      & $0.50$                                      \\ 
\hline
Arvaniti et. al (2018) \cite{Arvaniti2018AutomatedLearning}                          & $0.55$                                        & $0.49$  
\\ \hline

Pathologist $2$                                         & $0.65$                                      & $-$                                      \\ 
\hline
\hline
\multicolumn{3}{c}{Core-Level Scoring} \\
\hline
\hline
WeGleNet                                      & $0.76$                                      & $0.67$                                      \\ 
\hline
Arvaniti et. al (2018) \cite{Arvaniti2018AutomatedLearning}                          & $0.75$                                        & $0.71$                                        \\ \hline
Bulten et al. (2020) \cite{Bulten2020AutomatedStudy}                          & $0.72$                                        & $0.70$                                        \\ \hline
Pathologist $2$                                         & $0.71$                                      & $-$                                      \\ 
\hline

\end{tabular}
\end{table}

The obtained results are in line with our previous experiments, and WeGleNet performed comparably to the fully-supervised approach used by Arvaniti et. al \cite{Arvaniti2018AutomatedLearning}. We reached a better $\kappa$ value ($\kappa = 0.59$ against $\kappa = 0.53$) with the first pathologist, and similar performance was observed using the annotations from the second pathologist ($\kappa = 0.50$ against $\kappa = 0.49$). In addition, Figure \ref{fig:cm} showed that most of the errors were conducted between adjacent classes.

Regarding the core-level Gleason scoring, the performance was also similar to previous works in the test set. A $\kappa$ of $0.76$ and $0.67$ was obtained with each pathologist, respectively. In average, the obtained $\kappa$ ($0.715$) is similar to the one obtained by Arvaniti et al. ($0.730$) and Bulten et al. ($0.719$). These results are at the same level of inter-pathologist agreement ($k=0.710$). In addition, our approach obtained accurate localization heat-maps validated in Section \ref{ssec:exp3} without using pixel-level annotations during training. 

%% file: 06_Conclusions.tex
\section{Conclusions}
\label{sec:conclusions}

In this work, we have presented WeGleNet, a weakly-supervised trained architecture able to obtain semantic segmentation maps of Gleason grades in prostate histology images. The model is trained using just global-level labels, the Gleason score obtained from medical history, and it is capable of locating the local cancerous patterns in the tissue according to its grade. 

Our proposed architecture makes use of multi-class segmentation layers after the feature-extraction stage, and a global-aggregation of the pixel-level probabilities into one representative value per class. Then, the output of the non-cancerous class (background) was sliced to obtain the loss of the model during training. This strategy allows us to obtain complementary maps in the architecture, without requiring complex post-processing of the output. In the experimental stage, we compared different global-aggregation layers and regularization techniques to optimize the model performance in the validation cohort. The log-sum-exponential pooling (LSE) showed superior performance than other layers, thanks to its ability to adapt the model to the specific domain via the adjustable parameter $r$. Thus, we have achieved a Cohen’s quadratic kappa ($\kappa$) of $0.67$ for the Gleason grading of local patterns in the validation cohort at the pixel level. During this optimization stage, we have observed a high correlation between global and local-level figures of merit. Thus, optimizing the proposed architecture using just global-labels involves improving the local-level localization of cancerous patterns. Additionally, we have compared the model performance with state-of-the-art supervised methods for semantic segmentation of Gleason grades in the test cohort. The proposed WeGleNet architecture performed similarly to supervised methods, without requiring any kind of pixel-level annotations during the training stage, reaching a pixel-level $k$ of $0.61$ and an average f1-score of $0.58$. The performance for the core-level Gleason scoring was similar to previous works, and comparable to inter-pathologist agreement in the test cohort, reaching an average $\kappa$ of $0.715$. These promising results constitute a step forward in the literature of the analysis of prostate histology images and could avoid the tedious process of pixel-level generation of ground truth by expert pathologists.

Further research will focus on generalizing the proposed method to be trained using entire slices of biopsies digitized as whole slide images, whose larger size presents an added challenge in developing weakly-supervised methods for locating local cancerous patterns.